\documentclass[a4paper,11pt]{article}
\pdfoutput=1 

\usepackage{jheppub} 
\usepackage{tikz}
\usepackage{pgfplots}
\usetikzlibrary{calc,decorations.pathmorphing,shapes}
\usepackage{standalone}
\usepackage{physics}
\usepackage{subcaption}

\usepackage[T1]{fontenc} 

\title{The Action of Geometric Entropy in Topologically Massive Gravity}

\author{Molly Kaplan}

\affiliation{Department of Physics, University of California at Santa Barbara, \\ Santa Barbara, CA 93106, U.S.A}

\emailAdd{mekaplan@ucsb.edu}

\abstract{Due to the presence of a gravitational anomaly in topologically massive gravity (TMG), the geometric entropy is no longer simply the Hubeny-Rangamani-Takayanagi (HRT) area; instead, it is given by the HRT area plus an anomalous contribution. We study the action of this geometric entropy on the covariant phase space of classical solutions for TMG with matter fields whose action is algebraic in the metric. The result agrees precisely with the action of HRT area operators in Einstein-Hilbert gravity given in \href{https://arxiv.org/abs/2203.04270}{arXiv:2203.04270}, i.e., it is a boundary-condition-preserving kink transformation. Furthermore, we show our result to be consistent with direct computations of semiclassical commutators of geometric entropies in pure TMG spacetimes asymptotic to planar AdS, as computed in \href{https://arxiv.org/abs/2206.00027}{	arXiv:2206.00027}.}

\begin{document} 
\maketitle
\flushbottom

\section{Introduction}
\label{sec:intro}
The study of entanglement entropy has contributed crucially to progress across theoretical physics. For instance, entanglement entropy has played an integral part in understanding the nature of quantum field theories \cite{Calabrese_2004}, as well as understanding topological order in quantum many-body systems \cite{Kitaev_2006, Levin_2006}. Additionally, in holography, a fundamental outcome of the Anti-de Sitter/Conformal Field Theory (AdS/CFT) correspondence is the relation between entanglement entropies in the CFT and geometric entropies $\sigma$ of codimension-2 extremal surfaces in the AdS bulk. This relation is described by the Ryu-Takayanagi (RT) correspondence \cite{Ryu_2006, Ryu_2006_2}, or by its covariant generalization, the Hubeny-Rangamani-Takayanagi (HRT) correspondence \cite{Hubeny_2007}. In the limit where the bulk is described by Einstein-Hilbert gravity, the geometric entropy $\sigma$ is just $A/4G$ where $A$ is the area of the surface, though higher derivative terms in the action provide additional corrections to $\sigma$ \cite{Dong_2014}.

Given an HRT surface defined by a boundary region $R$, the area $A_{HRT}[R]$ of this HRT surface is hence of great interest (even without reference to its CFT dual). In particular, one can think of $A_{HRT}[R]$ as a quantum operator in the bulk by promoting it from its classical role as a function on the gravitational phase space. The action of this operator in semiclassical gravity was studied directly in \cite{Kaplan_2022}, where it was found to generate a boundary-condition-preserving kink transformation. As will be described in more detail in Section \ref{sec:HRTflow} below, this transformation acts as a relative boost between the entanglement wedges on either side of the HRT surface. Prior to the explicit study of the action of $A_{HRT}[R]$ in  \cite{Kaplan_2022}, there were many closely related results in various contexts \cite{Carlip_1995, Thiemann_1993, Kastrup_1994, Kuchar_1994, Donnelly_2016, Speranza_2018, Chandrasekaran_2019}, which suggested a similar form for the transformation. Most relevant to our work here are \cite{Jafferis_2014, Ceyhan_2018, Faulkner_2019, Bousso_2020, Bousso_2020_2}, which suggested that the HRT area action would generate this boost-like transformation based on comparison with modular Hamiltonians. These modular Hamiltonians are given by the expression $K=-\log\rho$ for some state $\rho$.

In \cite{Kaplan_2022} we determined the action of $A_{HRT}[R]$ in the gravitational phase space, working in AdS$_D$ Einstein-Hilbert gravity and including arbitrary minimally coupled matter. To understand the action on the phase space, the paper calculated Poisson brackets between $A_{HRT}[R]$ and certain gravitational data. Semiclassically, these Poisson brackets correspond to commutators, up to a factor of $i$. That paper also computed explicit Poisson brackets between HRT areas defined by different boundary regions $R$ in the Poincar\'e AdS groundstate for $2+1$ Einstein-Hilbert gravity. This calculation proceeded by starting with the boundary stress tensor algebra, then extending it to an area commutator via the Leibniz rule. The goal of the current work is to extend the results of \cite{Kaplan_2022} to $2+1$-dimensional asymptotically AdS spacetimes with a chiral boundary CFT, by which we mean a CFT with unequal left and right central charges. 

The present work is inspired by \cite{Zou_2022}, where the authors find an explicit expression for the modular commutator \cite{Kim_2022, Kim_2022_2} in $1+1$D chiral CFTs. This modular commutator is defined as $J(A,B,C)_{\rho}=\langle [K_{AB},K_{BC}] \rangle$, where $K_{AB}$ and $K_{BC}$ are the boundary modular Hamiltonians associated with regions $AB$ and $BC$, respectively, $\rho=\rho_{ABC}$ is some state, and $\langle ... \rangle$ denotes expectation values in that state. For contiguous CFT intervals $A$, $B$, and $C$ on a Cauchy surface $\Sigma$, the authors of \cite{Zou_2022} find a modular commutator given by
\begin{equation}\label{eq:Zou}
    J(A,B,C)_{\Omega} = \frac{\pi c_L}{6}(2\eta_v-1) - \frac{\pi c_R}{6}(2\eta_u-1) 
\end{equation}
where $c_L, c_R$ are the left and right central charge, respectively, $\ket{\Omega}$ is the vacuum state on $\Sigma$, and $u=t-x$ and $v=t+x$ are light cone coordinates. Additionally, $\eta_u=\frac{(u_1-u_2)(u_3-u_4)}{(u_1-u_3)(u_2-u_4)}$ and $\eta_v=\frac{(v_1-v_2)(v_3-v_4)}{(v_1-v_3)(v_2-v_4)}$, where $(u_1,v_1)$ and $(u_2,v_2)$ are the anchor points of region $A$, $(u_2,v_2)$ and $(u_3,v_3)$ are the anchor points of region $B$, and $(u_3,v_3)$ and $(u_4,v_4)$ are the anchor points of region $C$.

We would like to compare Eq.~\eqref{eq:Zou} to bulk area commutators for general pure states in the bulk. By the Jafferis-Lewkowycz-Maldacena-Suh (JLMS) relation \cite{Jafferis_2016}, we have $K_{R}=\frac{A_{ext}}{4G}+K_{bulk}+ S_{corrections}$, where $A_{ext}$ is the area of an extremal surface corresponding to the boundary region $R$, $K_{bulk}$ is the modular Hamiltonian of the bulk region enclosed by the extremal surface, and $S_{corrections}$ arise when computing quantum corrections. These include Wald-like terms and higher derivative corrections, allowing for terms built from extrinsic curvatures. In semiclassical gravity, we can safely ignore $K_{bulk}$, giving $\sigma[R] \approx K_R$, where $\sigma$ may include the higher derivative corrections found in $S_{corrections}$. However, this introduces a potential subtlety: $\sigma[R] \approx K_R$ is true in \textit{any} state, whereas the modular commutator is given by a commutator of vacuum modular Hamiltonians.

We can remedy this issue by noting that, in the bulk semiclassical approximation,
\begin{equation}
    e^{i\lambda \sigma}\ket{\psi} \approx e^{i\lambda K_{\psi}}\ket{\psi}
\end{equation}
for a modular Hamiltonian $K_{\psi}$ defined by the holographic pure state $\ket{\psi}$, and some arbitrary parameter $\lambda$\footnote{This approximation will be explained in detail in the forthcoming work \cite{Dong_Unpublished}.}. This is enough to compute expectation values of commutators. In particular, we find $\langle [ K_{AB}, K_{BC}]\rangle_{\Omega} = \langle [ \sigma[AB],  \sigma[BC] ] \rangle_{\Omega}$ for $K_{AB}$, $K_{BC}$ defined in the state $\ket{\Omega}$. In Einstein-Hilbert gravity, $\sigma$ is an HRT area. In this case, Eq.~\eqref{eq:Zou} reduces to $J(A,B,C)_{\Omega}=\frac{\pi c}{3}(\eta_v-\eta_u)$. This is exactly the area commutator computed in \cite{Kaplan_2022}.

We now wish to extend the derivation of area commutators to find agreement with the full modular commutator in Eq.~\eqref{eq:Zou}. To do this, we need to modify our bulk spacetime so that it is dual to a boundary CFT with $c_L \neq c_R$. This can be accomplished by adding to the Einstein-Hilbert action a gravitational Chern-Simons term, which is a higher-derivative term that preserves bulk diffeomorphism invariance, but which introduces a gravitational anomaly in the dual CFT. This anomaly manifests as either a non-conservation of the boundary stress tensor or, equivalently, as an anti-symmetric part of the boundary stress tensor, thus allowing for chiral behavior in our boundary CFT. This anomaly arises due to the theory's sensitivity to the choice of coordinate system at the boundary. In $2+1$ bulk dimensions, the resulting bulk theory is known as topologically massive gravity (TMG); see \cite{Deser_1982, Deser_1982_2, Deser_1991} for original references. Previous work studying TMG in a holographic context includes \cite{Kraus_2006, Solodukhin_2006, Hotta_2008, Skenderis_2009}.

In TMG, due to the presence of the bulk Chern-Simons term, the geometric entropy is no longer given by just the HRT area. Instead, the geometric entropy is given by the HRT area plus an extra term, as derived from the bulk perspective in \cite{Castro_2014} (using methods based on those in \cite{Lewkowycz_2013}). We will call this the TMG geometric entropy, and denote the corresponding quantum operator as $\sigma_{TMG}[R]$. We can gain more intuition about the TMG geometric entropy by comparing with Einstein-Hilbert gravity, where we can think of the HRT area as the action of a massive particle propagating in the bulk. In contrast, in TMG, the geometric entropy is given by the action of a massive spinning particle in the bulk. See \cite{Azeyanagi_2015, Iqbal_2016} for other studies on entanglement entropy in the presence of gravitational anomalies.

Using the TMG geometric entropy computed in \cite{Castro_2014}, we derive vacuum expectation values of commutators of $\sigma_{TMG}$, which indeed match the modular commutator in Eq.~\eqref{eq:Zou}. We also derive the Hamiltonian flow generated by $\sigma_{TMG}$ in semiclassical gravity. This direct calculation is a first step in understanding the action of geometric entropies in general higher-derivative gravitational theories. References \cite{Bousso_2020_2, Kaplan_2022} suggest that geometric entropy flow should remain a boundary-condition-preserving kink transformation, even with the inclusion of higher-derivative corrections to the Einstein-Hilbert action. This conjecture will be studied further in \cite{Dong_Unpublished_2}. Our work here is an explicit verification of this hypothesis for TMG.

In Section \ref{sec:HRTflow}, we reformulate the derivation of the phase space flow generated by HRT areas in the language of Peierls brackets \cite{Peierls_1952}, which are equivalent to the more familiar Poisson brackets but are more convenient for our purposes. We then use this same Peierls bracket method to compute TMG geometric entropy flow in Section \ref{sec:entropyFlowTMG}. The result is a boundary-condition-preserving kink transformation, which is exactly the transformation found for HRT area flow \cite{Kaplan_2022}. This result holds in spacetimes without matter. More generally, it holds to first order in the flow parameter for spacetimes with matter fields whose action is algebraic in the metric. This includes the usual two-derivative scalar, Yang-Mills, and Proca fields.

In Section \ref{sec:GEcomms}, we compute the algebra of TMG entropy operators. We use the bulk perspective throughout this calculation, taking special care to include the Chern-Simons contribution to the boundary stress tensor in Section \ref{sec:T}, and computing $\sigma_{TMG}$ for general states in Poincar\'e AdS$_3$ in Section \ref{sec:S}. In Section \ref{sec:algebra_flow}, we calculate the $\sigma_{TMG}$ algebra in the vacuum using TMG geometric entropy flow, and in Section \ref{sec:algebra_T} we calculate the $\sigma_{TMG}$ algebra in general states using the stress tensor algebra. We provide this calculation to make contact with \cite{Kaplan_2022}, and as an independent check on our main result in Section \ref{sec:entropyFlowTMG}. In Section \ref{sec:disjoint}, we extend the work of \cite{Zou_2022} by finding the TMG entropy algebra for disjoint boundary regions $A$, $B$, and $C$. Finally, in Section \ref{sec:discussion}, we conclude with some comments and possible future directions.

\section{Geometric entropy flow}
This section derives the geometric flow induced by the TMG geometric entropy $\sigma_{TMG}$. The result applies to asymptotically AdS$_3$ spacetimes with negative cosmological constant $\Lambda$ and without matter. It also holds to first order in the flow parameter $\lambda$ in spacetimes with matter fields whose action is algebraic in the metric, i.e., "standard matter". In order to quantify the entropy flow, we compute Peierls brackets between the geometric entropy and data on a particular Cauchy slice. In the bulk semiclassical approximation, Peierls brackets describe commutatation relations between operators (up to the usual factor of $i$). Importantly, a Peierls brakcet $\{A,B\}$ is only well-defined if both $A$ and $B$ are gauge-invariant.

In the dual CFT, we consider the entanglement entropy of an achronal region $R$. In semiclassical Einstein-Hilbert gravity, the associated geometric entropy is given by $1/4G$ times the area of the corresponding HRT surface, which is the minimal codimension-2 extremal surface anchored to $\partial R$ that satisfies the homology constraint of \cite{Headrick_2007}. As we will see, in TMG the geometric entropy is instead given by $1/4G$ times the area of some surface $\gamma$ (which lies in a Cauchy slice $\Sigma$), plus an additional term related to other data on $\Sigma$. The surface $\gamma$ is the one which extremizes the entropy functional $\sigma_{TMG}$. The surface $\gamma$ generally differs from the HRT surface one would find for $c_L=c_R$, but they are the same when matter is not present \cite{Castro_2014}.

Our Peierls bracket analysis will focus on the effect of geometric entropy flow on Cauchy data on $\Sigma$. One can then solve the equations of motion to find the action on the rest of the spacetime. In particular, we compute the bracket between the geometric entropy $\sigma$ and $K^{ij}$, the extrinsic curvature of the codimension-$1$ surface $\Sigma$. Readers unfamiliar with the Peierls bracket may wish to consult \cite{Harlow_2021} (and references therein) for background information.

The procedure to compute Peierls brackets starts by adding $\sigma$ as a source to the action. Then, we solve the new equations of motion to find the retarded and advanced solutions for the extrinsic curvature, denoted as $D^-K^{ij}_{tot}$ and $D^+K^{ij}_{tot}$, respectively. The rest of the data on $\Sigma$ remains unchanged. Finally, the desired Peierls bracket is defined by
\begin{equation}\label{eq:PBdef}
        \left\{\sigma,K^{ij}(x)\right\}=D^-K^{ij}_{tot}(x)-D^+K^{ij}_{tot}(x).
\end{equation}
Section \ref{sec:HRTflow} computes $\{A_{HRT}[R]/4G,K^{ij}(x)\}$ in Einstein-Hilbert gravity to illustrate the Peierls bracket method. We then compute $\{\sigma_{TMG}[R],K^{ij}(x)\}$ in Section \ref{sec:entropyFlowTMG}.

\subsection{Revisiting HRT area flow in semiclassical Einstein-Hilbert gravity}\label{sec:HRTflow}
In this section, we directly compute Peierls brackets in asymptotically AdS$_D$ Einstein-Hilbert gravity with standard matter. This commutator was previously computed in \cite{Kaplan_2022} using the canonical commutation relations of Einstein-Hilbert gravity; here, we instead use the ADM formalism \cite{Arnowitt_2008} and the Peierls bracket method. We perform this calculation as a simple illustration of this method, before applying it to the more complicated case of TMG. As we will show, our result here matches the previous result.

In the ADM formalism, we decompose the metric according to
\begin{equation}
    ds^2 = (-N^2+N_iN^i)dt^2 + 2N^idxdt + h_{ij}dx^idx^j,
\end{equation}
where $x^i$ are coordinates in a Cauchy slice $\Sigma$ and $h_{ij}$ is the induced metric on $\Sigma$. Using this decomposition, up to boundary terms the action can be written as \cite{Wald_1984}
\begin{equation}\label{eq:action}
\begin{split}
    I =& \int_{\mathcal{M}} dtd^{D-1}x \sqrt{-g} \left[ \frac{1}{16\pi G}(R - 2\Lambda) + \mathcal{L}_M \right] \\
    =&\int_{\mathcal{M}} dtd^{D-1}x N\sqrt{h}\left[\frac{1}{16\pi G}( r - K^2 + K_{ij}K^{ij}-2\Lambda) + \mathcal{L}_M\right],
\end{split}
\end{equation}
where $\mathcal{M}$ is the entire bulk manifold, $r$ is the Ricci scalar on $\Sigma$, $\mathcal{L}_M$ is the matter Lagrangian, and $K^{ij}$ is the extrinsic curvature on $\Sigma$. The extrinsic curvature is defined as 
\begin{equation}
    K_{ij}=\frac{1}{2N}(\dot h_{ij} - D_iN_j - D_j N_i),
\end{equation}
with $D_i$ the covariant derivative on $\Sigma$. We write the trace of $K_{ij}$ as $K=h_{ij}K^{ij}$.

Following the Peierls bracket method, we now add the geometric entropy defined by a boundary region $R$ as a source to the action. For semiclassical Einstein-Hilbert gravity, the geometric entropy is given by $1/4G$ times the area of the HRT surface $\gamma$ corresponding to the boundary region $R$. Additionally, we choose $\Sigma$ so that it contains $\gamma$. The HRT area is given by \cite{Kaplan_2022}
\begin{equation}
\begin{split}
    \frac{A_{HRT}[R]}{4G} =& \frac{1}{4G}\int_{\gamma}d^{D-2}w \sqrt{q(w)} \\
    =& \frac{1}{4G}\int_{\mathcal{M}}dtd^{D-1}x \sqrt{q(x)} \delta_{\Sigma}(\gamma,x)\delta(t-t_{\Sigma}),
\end{split}
\end{equation}
where $q_{AB}$ is the metric on the HRT surface, $\delta_{\Sigma}(\gamma,x)$ is a one-dimensional Dirac delta-function on the Cauchy slice which localizes $x$ to $\gamma$, and $t_{\Sigma}$ is the time associated with the Cauchy slice. Adding this to the action in Eq.~\eqref{eq:action} with (infinitesimal) weight $\lambda$ gives the modified action
\begin{equation}
\begin{split}
    I' = \int_{\mathcal{M}} dtd^{D-1}x \bigg(\frac{N}{16\pi G}& \sqrt{h(x,t)}[ r(x,t) - K^2(x,t) + K_{ij}(x,t)K^{ij}(x,t)-2\Lambda] \\
    &+ N\sqrt{h(x,t)} \mathcal{L}_M(x,t) + \frac{\lambda}{4G} \sqrt{q(x)} \delta_{\Sigma}(\gamma,x)\delta(t-t_{\Sigma})\bigg).
\end{split}
\end{equation}

Next, we set $\delta I'=0$ and solve the resulting equations of motion. The modification of the action introduces a new term containing $\delta(t-t_\Sigma)$, and so, to cancel this term in the equation of motion, we need another term proportional $\delta(t-t_\Sigma)$. As we will show, this can be achieved with an ansatz in which advanced and retarded solutions of the induced metric remain continuous but in which advanced and retarded solutions for $K^{ij}$ involve terms proportional to a Heaviside-function $\Theta(t-t_\Sigma)$. We will denote the retarded solution by $D^-K^{ij}_{tot}$ and the advanced solution by $D^+K^{ij}_{tot}$. Below, we will focus only on "relevant" terms in the equation of motion. This simply means we will only keep terms proportional to $\delta(t-t_\Sigma)$ which, with the above ansatz, are simply those containing time-derivatives of $D^{\pm}K^{ij}_{tot}$.

To find $\delta I'$, we need to understand the functional derivatives of $h_{ij}$ and $q_{ij}$ with respect to $h_{ij}$. These are given \cite{Kaplan_2022} by 
\begin{eqnarray}\label{eq:delhdelh}
     &\frac{\delta h_{kl}(x)}{\delta h_{ij}(y)} = \delta_k^i\delta_l^j \delta^{(D-1)}(x-y) \\
     \label{eq:delqdelh}
     &\frac{\delta q_{AB}(x)}{\delta h_{ij}(y)} = \frac{\partial y^i}{\partial \tilde x^A}\frac{\partial y^j}{\partial \tilde x^B}\delta_{\gamma}^{(D-2)}(x,\tilde x(y))\delta_{\Sigma}(\gamma,y).
\end{eqnarray}
We also need to understand the variation of $K_{ij}$ with respect to $h_{ij}$, which is
\begin{equation}
    \frac{\delta K_{kl}(x)}{\delta h_{ij}(y)} = \frac{1}{2N}\partial_t \bigg( \frac{\delta h_{kl}(x)}{\delta h_{ij}(y)}\bigg),
\end{equation}
and which can be evaluated fully using Eq.~\eqref{eq:delhdelh}. Finally, we need the variation of $A_{HRT}[R]$ with respect to $h_{ij}$, which is given by
\begin{equation}\label{eq:varA}
    \frac{\delta A_{HRT}[R]}{\delta h_{ij}(y)} = \frac{1}{2}\int_{\mathcal{M}}dtd^{D-1}x  \frac{\delta q_{AB}(x)}{\delta h_{ij}(y)}q^{AB}(x) \delta_{\Sigma}(\gamma,x)\delta(t-t_{\Sigma}),
\end{equation}
where, when the metric and $K_{ij}$ are evaluated at $t_\Sigma$, we do not write their explicit $t$-dependence. We will evaluate this fully by inserting Eq.~\eqref{eq:delqdelh}. This provides all of the pieces needed to evaluate the variation of the modified action.

Keeping only the relevant terms, we have
\begin{equation}\label{eq:delI}
\begin{split}
    \delta I_{rel}'=&  \frac{1}{16\pi G} \int dt \sqrt{h(t,y)} \delta h_{ij}(t,y) \bigg(\partial_t[K(t,y)h^{ij}(t,y)-K^{ij}(t,y)] \\
    &+ 2\pi\lambda \frac{\sqrt{q(\tilde x(y))}}{\sqrt{h(y)}}q^{AB}(\tilde x(y))\frac{\partial y^i}{\partial \tilde x^A}\frac{\partial y^j}{\partial \tilde x^B} \delta(t-t_{\Sigma})\delta_{\Sigma}(\gamma,y)\bigg).
\end{split}
\end{equation}
Notice that $\mathcal{L}_M$ does not factor into our calculation as long as it does not contain any extrinsic curvature components. This is true for standard matter, as defined above.
We will now solve for the effect of the new source term at first order in $\lambda$ about a background solution of the $\lambda=0$ theory.  For source strength $\lambda$, we write the extrinsic curvature at this order in the form
\begin{equation}\label{eq:Kdef}
    K^{ij}(t,y) = \tilde{K}^{ij}(t,y)+ \lambda D^{\pm}K_{tot}^{ij}(t,y),
\end{equation} 
where $\tilde{K}^{ij}$ is the original  extrinsic curvature of the $\lambda =0$ background.\footnote{The calculation of HRT area flow is unaffected by the smoothness of $\tilde K^{ij}$, which is important for including matter in the background spacetime. This is explained in more detail at the end of this section.} As discussed above, $D^{\pm} K^{ij}_{tot}$ must contain terms with $\Theta(t-t_\Sigma)$, but it can also contain continuous terms. Thus, we can write $D^{\pm}K^{ij}_{tot}=D^{\pm}K^{ij}+D^{\pm}K^{ij}_{cont}$, where $D^{\pm}K^{ij}$ contains all Heaviside-function terms and $D^{\pm}K^{ij}_{cont}$ contains all continuous terms. Since the continuous terms in the advanced and retarded solutions must agree on $\Sigma$, their difference vanishes in the Peierls bracket, and we can rewrite Eq.~\eqref{eq:PBdef} as
\begin{equation}\label{eq:PBdefdisc}
    \{A_{HRT}[R],K^{ij}(t_\Sigma, y)\}=D^-K^{ij}(t_\Sigma, y)-D^+K^{ij}(t_\Sigma, y).
\end{equation}
We now wish to solve for the advanced and retarded solutions to evaluate this Peierls bracket.

Plugging Eq.~\eqref{eq:Kdef} into Eq.~\eqref{eq:delI} and setting $\delta I_{rel}'=0$, we have
\begin{equation}
\begin{split}
    \partial_t D^{\pm}K^{ij}(t,y)-h_{kl}(t,y)&\partial_tD^{\pm}K^{kl}(t,y)h^{ij}(t,y) = \\
    &2\pi \frac{\sqrt{q(\tilde x(y))}}{\sqrt{h(y)}}q^{AB}(\tilde x(y))\frac{\partial y^i}{\partial \tilde x^A}\frac{\partial y^j}{\partial \tilde x^B} \delta_{\Sigma}(\gamma,y)\delta(t-t_\Sigma).
\end{split}
\end{equation}
Integrating over time\footnote{Because the induced metric $h_{ij}$ depends on time, our expressions for $D^{\pm}K^{ij}$ are not the exact results of these integrals. Instead, Eq.~\eqref{eq:retDHRT} and \eqref{eq:advDHRT} give only the discontinuous terms. If we expand each metric as a power series in $t$, then our expressions for $D^{\pm}K^{ij}$ come from considering only the term proportional to $\theta(t-t_\Sigma)$; higher-order terms in the metric give continuous terms which do not contribute to the Peierls bracket.} and performing a trace reverse gives our two solutions
\begin{eqnarray}\label{eq:retDHRT}
    D^-K^{ij}(t,y)=& 2\pi \frac{\sqrt{q(\tilde x(y))}}{\sqrt{h(y)}}\delta_{\Sigma}(\gamma,y)\Theta(t-t_\Sigma)\bigg(q^{AB}(\tilde x(y))\frac{\partial y^i}{\partial \tilde x^A}\frac{\partial y^j}{\partial \tilde x^B}-h^{ij}(y)\bigg) \\
    \label{eq:advDHRT} D^+K^{ij}(t,y)=& -2\pi \frac{\sqrt{q(\tilde x(y))}}{\sqrt{h(y)}}\delta_{\Sigma}(\gamma,y)\Theta(t_\Sigma-t)\bigg(q^{AB}(\tilde x(y))\frac{\partial y^i}{\partial \tilde x^A}\frac{\partial y^j}{\partial \tilde x^B}- h^{ij}(y)\bigg).
\end{eqnarray}
Finally, using the Peierls bracket definition in Eq.\eqref{eq:PBdef}, we arrive at our result
\begin{equation}\label{eq:PB_EH}
\begin{split}
    \bigg\{\frac{A_{HRT}[R]}{4G},K^{ij}(t_\Sigma,y)\bigg\} =& 2\pi \frac{\sqrt{q(\tilde x(y))}}{\sqrt{h(y)}}\delta_{\Sigma}(\gamma,y)\bigg(q^{AB}(\tilde x(y))\frac{\partial y^i}{\partial \tilde x^A}\frac{\partial y^j}{\partial \tilde x^B} - h^{ij}(y)\bigg) \\
    =& -2\pi \hat \delta_{\Sigma}(\gamma,y)\perp^i\perp^j,
\end{split}
\end{equation}
where $\perp^i$ is the unit normal to $\gamma$ in $\Sigma$, and $\hat{\delta}_{\Sigma}(\gamma,y)=\frac{\sqrt{q(\tilde{x}(y))}}{\sqrt{h(y)}} \delta_{\Sigma}(\gamma,y)$ is a one-dimensional Dirac delta-function of the proper distance between $x$ and $\gamma$ measured along geodesics in $\Sigma$ orthogonal to $\gamma$. Since $h^{ij}$ remains unchanged under the addition of the source term $\sigma$, the Peierls bracket $\{A_{HRT}[R],h^{ij}(t_\Sigma,y)\}$ vanishes. The flow thus adds a $\delta$-function (times $-2\pi\lambda$) to $K^{\perp\perp}$, but leaves all other initial data on $\Sigma$ unchanged. This precisely matches our previous result for the HRT area flow in \cite{Kaplan_2022}, which we arrived at using the standard Poisson brackets of phase space variables on the Cauchy slice. 

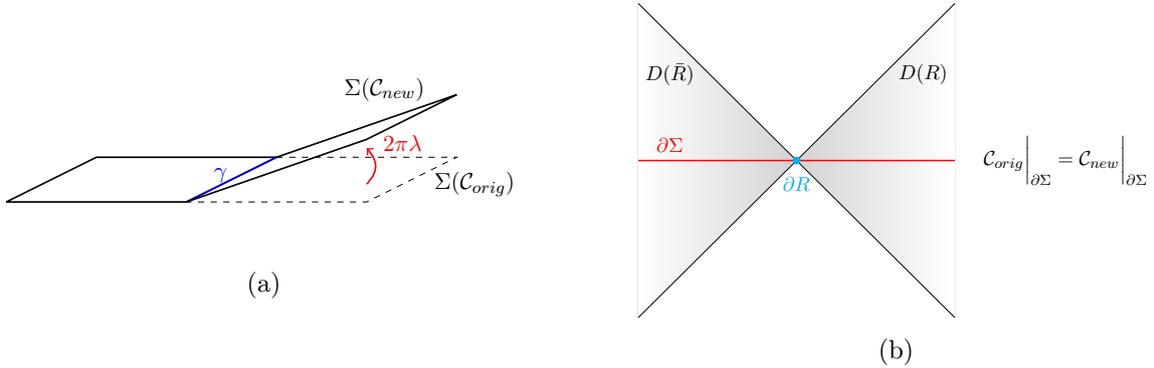
\begin{figure}
    \centering
    \begin{subfigure}[][][c]{0.45\textwidth}
        \centering
        \begin{tikzpicture}[scale=3]
    \draw[dashed] (.5,0) -- (1.5,0) -- (1,-.25) -- (0,-.25);
    \draw[thick] (-1,-.25) -- (-.5,0) -- (.5,0) -- (0,-.25) -- (-1,-.25);
    \draw[blue, thick] (0,-.25) -- (.5,0);
    \draw[thick] (0,-.25) -- (1,.1) -- (1.5,.35) -- (.5,0);
    \draw[red, ->, thick] plot [smooth, tension=1] coordinates { (1,-.15)(1.05,-.05)(1,.05)};
    \draw[red] (1.2,.08) node {$2\pi\lambda$};
    \draw[blue] (.19,-.1) node {$\gamma$};
    \draw (1.6,-.15) node {$\Sigma(\mathcal{C}_{orig})$};
    \draw (1.1,.38) node {$\Sigma(\mathcal{C}_{new})$};
    \vspace{2cm};
\end{tikzpicture}
        \vspace{.1cm}
        \caption{}
        \label{fig:bulkflow}
    \end{subfigure}
    \hfill
    \begin{subfigure}[][][c]{0.45\textwidth}
        \centering
        \begin{tikzpicture}[scale=3]
    \fill[left color=white,right color=lightgray!70] (-1,-1) -- (0,0) -- (-1,1);
    \fill[right color=white,left color=lightgray!70] (1,1) -- (0,0) -- (1,-1);
    \draw (-1,1) -- (1,-1);
    \draw (-1,-1) -- (1,1);
    \draw[red,thick] (-1,0) -- (1,0);
    \draw[red] (-.8,.1)  node {$\partial \Sigma$};
    \fill[cyan] (0,0) circle (.7pt);
    \draw[cyan] (0,-.15) node {$\partial R$};
    \draw (-.8,.55) node {$D(\bar{R})$};
    \draw (.8,.55) node {$D(R)$};

    \draw (1.7,0) node {$\mathcal{C}_{orig}\bigg|_{\partial \Sigma} = \mathcal{C}_{new}\bigg|_{\partial \Sigma}$};
\end{tikzpicture}
        \caption{}
        \label{fig:bndyflow}
    \end{subfigure}
    \caption{The geometry of the boundary-condition-preserving kink transformation. Figure \ref{fig:bulkflow} shows the transformation of the Cauchy data on $\Sigma$ in the bulk, from $\mathcal{C}_{orig}$ to $\mathcal{C}_{new}$. The flow induces a relative boost with parameter $2\pi\lambda$ between the left and right sides of $\Sigma$. Figure \ref{fig:bndyflow} depicts the transformation induced in the boundary, showing the domains of dependence $D(R)$ of $R$ and $D(\bar{R})$ of $\bar{R}$. $\partial R$ is the intersection between the boundary and $\gamma$, and $\partial\Sigma$ is the boundary of a smooth bulk Cauchy surface $\Sigma$ in the original spacetime. On the surface $\partial \Sigma$, $\mathcal{C}_{orig}=\mathcal{C}_{new}$. All boundary observables on that surface are preserved by the flow generated by $A_{HRT}[R]$.}
    \label{fig:flow}
\end{figure}

How can we understand this result geometrically? We can integrate Eq.~\eqref{eq:PB_EH} to yield the effect of a finite flow by a parameter $\lambda$, and we see that the flow induced by the HRT area introduces a relative boost in $\Sigma$, on either side of $\gamma$. This "kinks" the data on the Cauchy slice in the bulk, as shown in Figure \ref{fig:bulkflow}, and the rest of the solution is determined by the equations of motion. However, we must take special care with boundary conditions. In particular, since $\Sigma$ represents a definite instant of time, the boundary of $\Sigma$ ($\partial \Sigma$) must remain fixed due to the asymptotic AdS boundary conditions. This is shown in Figure \ref{fig:bndyflow}. Following \cite{Kaplan_2022}, we refer to this transformation as a boundary-condition-preserving kink transformation.

The treatment of boundary conditions is in contrast to the original kink transformation introduced in \cite{Bousso_2020_2}. Defining $\mathcal{K}[\gamma]$ as the generator of this original kink transformation by $\lambda$, then $\mathcal{K}[\gamma]$ has the same bulk action as $A_{HRT}[R]/4G$ but has different boundary conditions as $\mathcal{K}[\gamma]$ would instead introduce a relative boost on either side of $\partial R$, the boundary of $\gamma$. In particular, this $\mathcal{K}[\gamma]$ also acts as a relative boost at the boundary. Then, defining $H_R$ to be ($2\pi$ times) the generator of the boundary one-sided boost (taken to generate flow toward the future in the right wedge), the relation can be expressed in the form $\frac{A_{HRT}[R]}{4G} = H_R + \mathcal{K}[\gamma]$. This notation will be useful in Section \ref{sec:algebra_flow}.

As a final comment, we note that integrating the flow to finite $\lambda$ requires an understanding of the Peierls Bracket evaluated at certain background solutions that are non-smooth as well as at those that are smooth. This is because, if we take a smooth background solution and apply HRT area flow, then the solution immediately becomes non-smooth due to the kink transformation. In the analysis above, we allowed for non-smooth $\tilde{K}^{ij}$, so there is no obstruction to integrating to finite $\lambda$. However, as we will see, integrating to finite $\lambda$ is more difficult for TMG entropy flow.

\subsection{Entropy flow in TMG}\label{sec:entropyFlowTMG}
We now apply the Peierls bracket method to TMG in spacetimes asymptotic to AdS$_3$ with standard matter. In this theory, the bulk action is
\begin{equation}
    I = I_{EH}-\beta I_{CS},
\end{equation}
where $I_{EH}$ is the Einstein-Hilbert action and $I_{CS}$ is the Chern-Simons action, defined as
\begin{equation}
    I_{CS}= \int_{\mathcal{M}} \text{Tr}\bigg[\Gamma d\Gamma + \frac{2}{3}\Gamma^3 \bigg].
\end{equation}
The constant $\beta$ measures the anomaly coefficient. It is defined as
\begin{equation}
    \beta = \frac{c_L-c_R}{96\pi},
\end{equation}
and we can use this to write the left and right central charges as $c_L=c_0 + 48\pi \beta$ and $c_R=c_0 - 48\pi \beta$, where $c_0=3/2G$ is the central charge in the absence of the Chern-Simons term. For more information on the notation used above, we refer the reader to, e.g., \cite{Kraus_2006}.

Again, we have a boundary region $R$ and its corresponding HRT surface $\gamma$. We are interested in calculating the Peierls bracket between $\sigma_{TMG}[R]$, the geometric entropy determined by $R$, and $K^{ij}$, the extrinsic curvature of a Cauchy slice $\Sigma$ containing $\gamma$. To do so, we add $\sigma_{TMG}$ as a source to the action, take the variation, and solve the equations of motion. Thus, we are interested in solutions to the equation
\begin{equation}\label{eq:EOM}
    \delta I_{EH} - \beta \delta I_{CS} + \lambda \delta \sigma_{TMG}[R] = 0.
\end{equation}
In Section ~\ref{sec:HRTflow}, we calculated the relevant terms of $\delta I_{EH}$. As we will show, these terms will again be the only relevant terms in our TMG calculation. We are left with computing the relevant terms in $\delta I_{CS}$ and $\delta \sigma_{TMG}[R]$.

In what follows, we will fix our gauge so that $N=1$ and $N^i=0$. As we mentioned above, this gauge-fixing is allowed because a Peierls bracket $\{A,B\}$ must have gauge-invariant $A$ and $B$. 

\subsubsection{Calculating $\delta \sigma_{TMG}[R]$}
In TMG, the geometric entropy is modified by an additional term, and so is no longer given by the HRT area. Instead, by Equation (3.26) of \cite{Castro_2014}, the TMG geometric entropy defined by a boundary region $R$ is given by
\begin{equation}\label{eq:S}
    \sigma_{TMG}[R] = \frac{1}{4G} \int_\gamma ds \bigg(\sqrt{g_{\mu\nu}\dot X^{\mu} \dot X^{\nu}} - 32\pi G \beta \tilde \nu \cdot \nabla \nu \bigg),
\end{equation}
where $\gamma$ is the curve in spacetime that extremizes $\sigma_{TMG}$. The first term in the expression gives the area of $\gamma$. Without matter, $\gamma$ is the HRT surface one would find for $c_L=c_R$, and so the first term is the usual HRT area term. However, with matter present, $\gamma$ generally differs from the HRT surface. At each point of $\gamma$ the vectors $\nu^\mu$, $\tilde \nu^\mu$ define an orthonormal frame in the orthogonal plane. Now, the exact Poincar\'e AdS$_3$ solution has the metric
\begin{equation}\label{eq:metric}
    ds^2 = \frac{1}{z^2}(-dt^2+dx^2+dz^2),
\end{equation}
where we set $l_{AdS}=1$. Our spacetime is asymptotically AdS$_3$, so we can take our metric to asymptote to Eq.~\eqref{eq:metric} and use the corresponding coordinates $(t,x)$ to specify vectors on the boundary. We define the normal frame at the boundary as $\nu_{\partial} = \partial_t$ and $\tilde \nu_\partial= \partial_x$. We choose a Cauchy slice $\Sigma$ so that it contains $\gamma$. The surface $\gamma$ has two endpoints, and, in general, they do not have the same $t$-coordinates but they do lie on the same spacelike line on the boundary. We are free to choose $\Sigma$ to asymptote to that line, which is a boost of the constant $t$ slice by some boost parameter $\alpha$. Then, we can define the boundary vectors,
\begin{eqnarray}
    n_\partial^\mu=&(\cosh\alpha,-\sinh\alpha,0) \\
    \perp_\partial^\mu =& (-\sinh\alpha,\cosh\alpha,0)
\end{eqnarray}
where $n^\mu$ is the vector normal to $\Sigma$ and $\perp^\mu$ is the vector normal to $\gamma$ in $\Sigma$, and $n^\mu_\partial$ and $\perp_\partial^\mu$ are the boundary values of these vectors. Thus, in the bulk, we must have
\begin{eqnarray}
    \nu_\mu  =& \cosh\alpha (n_\mu + \tanh\alpha \perp_\mu) \\
    \tilde \nu^\mu =& \cosh\alpha (\tanh\alpha \,n^\mu +\perp^\mu).
\end{eqnarray}
Plugging into the last term in the action, we have
\begin{equation}
\begin{split}
    \tilde\nu^\mu \nabla \nu_\mu =& \perp^\mu v^\sigma K_{\mu\sigma} \\
    =& \perp^i v^j K_{ij},
\end{split}
\end{equation}
where, since $\perp^{\mu}$ and $v^{\mu}$ lie within $\Sigma$, we denote them with Latin indices.

Let us define the area of a surface $\xi$ as $A[\xi,g]$, where $g$ is the spacetime metric. Using our above result in Eq.~\eqref{eq:S}, we get
\begin{equation}
    \begin{split}
        \sigma_{TMG}[R] =& \frac{A[\gamma,g]}{4G} - 8\pi\beta \int_\gamma ds \perp^k v^l K_{kl}(s) \\
        =&  \frac{A[\gamma,g]}{4G} -8\pi\beta\int_\gamma dx \sqrt{q(x)} \perp^k v^l K_{kl}(x).
    \end{split}
\end{equation}
We now wish to vary the geometric entropy. In general, there are two components to this variation: the variation with respect to the surface and the variation with respect to $g$. However, because $\gamma$ extremizes $\sigma_{TMG}$, the variation with respect to the surface vanishes when evaluated at $\gamma$. So, we need only consider variations with respect to the metric, and can treat $\gamma$ as fixed.\footnote{For a more detailed argument on why we treat $\gamma$ as fixed, we refer the reader to the discussion at the start of Section 2.1 in \cite{Kaplan_2022}.} This yields
\begin{equation}\label{eq:Svariation}
\begin{split}
    \delta \sigma_{TMG}[R] =& \frac{\tilde{\delta} A[\gamma,g]}{4G} - 8\pi\beta \int_\gamma dx \sqrt{q(x)}\bigg[\perp^k v^l \delta K_{kl}(x) + \delta(\perp^k v^l) K_{kl}(x) \\
    &+ \frac{\delta \sqrt{q(x)}}{\sqrt{q(x)}} \perp^k v^l K_{kl}(x)\bigg] \\
    =& \frac{\tilde{\delta} A[\gamma,g]}{4G} - 8\pi\beta \int_\mathcal{M} dtd^2x \sqrt{q(x)}\bigg[\perp^k v^l \delta K_{kl}(x) + \delta(\perp^k v^l) K_{kl}(x) \\
    &+ \frac{\delta \sqrt{q(x)}}{\sqrt{q(x)}} \perp^k v^l K_{kl}(x)\bigg]\delta_{\Sigma}(\gamma,x)\delta(t-t_{\Sigma}),
\end{split}
\end{equation}
where $\tilde{\delta}A[\gamma,g]$ is the variation of the area at fixed $\gamma$, as given by Eq.~\eqref{eq:varA}.\footnote{While Eq.~\eqref{eq:varA} was defined as the HRT area variation, it holds more generally as the variation of an area $A[\xi,g]$ with respect to the metric.} As before, when the metric and $K_{ij}$ are evaluated at $t_\Sigma$, we do not write their explicit $t$-dependence. We also have expressions for $\delta K^{ij}$ and $\delta q_{ij}$, so all that is left to understand is the variation of $\perp^kv^l$. We have
\begin{equation}
    0=\delta(h_{ij}\perp^i\perp^j)=\perp^i\perp^j \delta h_{ij} + 2\perp_i\delta \perp^i,
\end{equation}
which gives
\begin{equation}
    \delta \perp^i = - \frac{1}{2}\perp^i\perp^k\perp^l\delta h_{kl} + \zeta v^i,
\end{equation}
where $\zeta$ is an unknown constant. Similarly, we can write
\begin{equation}
    0=\delta(h_{ij}v^iv^j)=v^iv^j \delta h_{ij} + 2v_i\delta v^i,
\end{equation}
which gives
\begin{equation}
    \delta v^i = - \frac{1}{2}v^iv^kv^l\delta h_{kl} + \eta \perp^i,
\end{equation}
where $\eta$ is a constant. Since we can treat $\gamma$ as a fixed surface, the only components of $\delta v^i$ we need are those which keep it normalized. Hence, $\eta=0$, so
\begin{equation}\label{eq:varv}
    \delta v^i = - \frac{1}{2}v^iv^kv^l\delta h_{kl}.
\end{equation}
To solve for $\zeta$, we now use
\begin{equation}
    0=\delta(h_{ij} \perp^i v^j)= \perp^i v^j\delta h_{ij} + \zeta,
\end{equation}
yielding $\zeta=-\perp^i v^j\delta h_{ij}$. Thus, we have
\begin{equation}\label{eq:varperp}
    \delta \perp^i = - \frac{1}{2}\perp^i\perp^k\perp^l\delta h_{kl} -v^i\perp^k v^l\delta h_{kl}.
\end{equation}

We can now use Eq.~\eqref{eq:varA} for the area variation, Eq.~\eqref{eq:varv} for the variation of $v^i$, and Eq.~\eqref{eq:varperp} for the variation of $\perp^i$ in Eq.~\eqref{eq:Svariation}. This gives
\begin{equation}\label{eq:delS}
\begin{split}
   \delta \sigma_{TMG}[R] =& \int dt \sqrt{q(\tilde{x}(y))} \bigg(- 4\pi\beta\perp^{i}  v^{j} \delta'(t-t_{\Sigma}) \\
    &+ \frac{1}{8G} q^{AB}(\tilde x(y))\frac{\partial y^i}{\partial \tilde x^A}\frac{\partial y^j}{\partial \tilde x^B}\delta(t-t_{\Sigma})-8\pi\beta v^k  v^l K_{kl}(y) \perp^i v^j \delta(t-t_{\Sigma})  \\
    &-4\pi\beta \perp^k  v^l K_{kl}(y) \perp^i \perp^j \delta(t-t_{\Sigma}) -4\pi\beta \perp^k  v^l K_{kl}(y) v^i v^j \delta(t-t_{\Sigma}) \\
    &+ 4\pi\beta q^{AB}(\tilde x(y))\frac{\partial y^i}{\partial \tilde x^A}\frac{\partial y^j}{\partial \tilde x^B}\perp^k v^l K_{kl}(y)\delta(t-t_{\Sigma}) \bigg)\delta_{\Sigma}(\gamma,y) \delta h_{ij}(y).
\end{split}
\end{equation}
As in Einstein-Hilbert gravity, the source variation includes terms proportional to a Dirac delta function, $\delta(t-t_{\Sigma})$. However, unlike the previous case, the first term in Eq.~\eqref{eq:delS} contains a time derivative of a delta function, denoted as $\delta'(t-t_\Sigma)$. We thus need other terms in the equation of motion to be proportional to $\delta$-functions and $\delta$-function derivatives, so they can cancel these new terms. As we will show in Section \ref{sec:delICS}, there will be terms in the equation of motion containing $\partial_t^2 D^{\pm}K^{ij}_{tot}$, so we can use the same ansatz as before. Namely, we will choose the advanced and retarded solutions of the induced metric, $D^{\pm}h^{ij}$, to be continuous, but choose solutions of the extrinsic curvature, $D^{\pm}K^{ij}_{tot}$, to have discontinuous terms proportional to a Heaviside-function.

As before, we will focus below only on "relevant" terms in the equation of motion. In this case, the relevant terms are ones containing $\partial_t D^{\pm}K^{ij}_{tot}$ or $\partial_t^2 D^{\pm}K^{ij}_{tot}$; with our ansatz, these will give the necessary $\delta(t-t_\Sigma)$ and $\delta'(t-t_\Sigma)$ terms.

\subsubsection{Calculating $\delta I_{CS}$}\label{sec:delICS}
The variation of $I_{CS}$, as given in \cite{Solodukhin_2006}, is
\begin{equation}\label{eq:delICS}
    \delta I_{CS} =  -2 \int_{\mathcal{M}} dtd^2x \sqrt{h} C^{\mu\nu} \delta g_{\mu\nu},
\end{equation}
where $C^{\mu\nu}$ is the Cotton tensor. In 3-dimensions, the Cotton tensor is
\begin{equation}
    \begin{split}
        C^{\mu\nu}=&\epsilon^{\mu\alpha\sigma} \nabla_{\alpha}\bigg(R^{\nu}_{\sigma} - \frac{1}{4}\delta^{\nu}_{\sigma}R\bigg) \\
        =& \epsilon^{\mu\alpha\sigma} \partial_{\alpha}R^{\nu}_{\sigma} + \epsilon^{\mu\alpha\sigma} \Gamma_{\alpha\lambda}^\nu R^{\lambda}_{\sigma} - \epsilon^{\mu\alpha\sigma} \Gamma_{\alpha\sigma}^\lambda R^{\nu}_{\lambda} -\frac{1}{4} \epsilon^{\mu\alpha\nu} \nabla_{\alpha} R.
    \end{split}
\end{equation}
It is of course important that all equations of motion are satisfied. However, for the purposes of this calculation, we need only consider the equations involving the $\mu=i$, $\nu=j$ components of the Cotton tensor, where $i,j$ are spatial indices. This is because variations of $\sigma_{TMG}[R]$ depend only on the induced metric, and not on any other metric components. Thus, these are the parts of the equation of motion changed by the introduction of the source, and all the other equations are constraints. The Bianchi identities guarantee that if the constraints are satisfied on any surface (e.g., to the past in a retarded solution) and if the equations of motion studied in this section are satisfied, the constraints will continue to hold on any surface. As a result, we need not explicitly check that the constraints are satisfied, and we may focus our attention on the remaining equations of motion.

After some cancellations, the spatial components of the Cotton tensor are
\begin{equation}\label{eq:Cij}
        C^{ij}=\frac{1}{2}\bigg(\epsilon^{i\alpha\sigma} \partial_{\alpha}R^{j}_{\sigma} + \epsilon^{i\alpha\sigma} \Gamma_{\alpha\lambda}^j R^{\lambda}_{\sigma} + \epsilon^{j\alpha\sigma} \partial_{\alpha}R^{i}_{\sigma} + \epsilon^{j\alpha\sigma} \Gamma_{\alpha\lambda}^i R^{\lambda}_{\sigma} \bigg),
\end{equation}
where we made explicit the symmetry under exchange of $i$ and $j$. As above, we are only interested in contributions containing a time derivative of the extrinsic curvature. Evaluating Eq.~\eqref{eq:Cij} in the gauge $N=1$ and $N^i=0$, and keeping only relevant terms, we arrive at
\begin{equation}
\begin{split}
    C^{ij}_{rel}=&\frac{1}{2}\epsilon^{it k}\bigg( \partial^2_t K^j_k - 4 \partial_t(K_{lk}K^{lj}) + \partial_t(K K^j_k) + K_l^j \partial_t K^l_k - K_{k}^j \partial_{t}K \bigg) \\
    &+ \frac{1}{2}\epsilon^{jt k}\bigg( \partial^2_t K^i_k - 4 \partial_t(K_{lk}K^{li}) + \partial_t(K K^i_k) + K_l^i \partial_t K^l_k - K_{k}^i \partial_{t}K\bigg).
\end{split}
\end{equation}
Using Eq.~\eqref{eq:delhdelh} and the identity
\begin{equation}
    \delta h_{ij}(x) = \frac{\delta h_{ij}(x)}{\delta h_{kl}(y)} \delta h_{kl}(y),
\end{equation}
we plug into Eq.~\eqref{eq:delICS}, yielding
\begin{equation}
\begin{split}
    \delta I_{CS, rel}=&-\int dt \sqrt{h(t,y)} \epsilon^{it k}\bigg( \partial^2_t K^j_k(t,y) - 4 \partial_t(K_{lk}(t,y)K^{lj}(t,y)) \\
    &+ \partial_t(K(t,y) K^j_k(t,y)) + K_l^j(t,y) \partial_t K^l_k(t,y) \\
    &- K_{k}^j(t,y) \partial_{t}K(t,y) \bigg)\delta h_{ij}(t,y) \\
    &-\int dt \sqrt{h(t,y)}\epsilon^{jt k}\bigg( \partial^2_t K^i_k(t,y) - 4 \partial_t(K_{lk}(t,y)K^{li}(t,y)) \\
    &+ \partial_t(K(t,y) K^i_k(t,y)) + K_l^i(t,y) \partial_t K^l_k(t,y)\\
    &- K_{k}^i(t,y) \partial_{t}K(t,y) \bigg) \delta h_{ij}(t,y).
\end{split}
\end{equation}
We now have equations for the relevant terms in $\delta I_{CS}$, $\delta \sigma_{TMG}[R]$ from Eq.~\eqref{eq:delS}, and $\delta I_{EH}$ from Eq.~\eqref{eq:delI}. In the next section, we combine these equations together to solve the equation of motion.

\subsubsection{Solving the modified equations of motion}\label{sec:TMGsolving}
To find the Peierls bracket, we start by using our expressions for $\delta I_{EH}$, $\delta \sigma_{TMG}[R]$, and $\delta I_{CS}$ in Eq.~\eqref{eq:EOM}. As in Section \ref{sec:HRTflow}, we write $K^{ij}=\tilde{K}^{ij}+\lambda D^{\pm}K^{ij}_{tot}$, where $\tilde{K}^{ij}$ is the extrinsic curvature before introducing the source $\lambda \sigma_{TMG}[R]$. As stated above, the modified extrinsic curvature, $D^{\pm}K^{ij}_{tot}$, has terms proportional to a Heaviside-function. It can also have terms containing $(t-t_\Sigma)\Theta(t-t_\Sigma)$ (i.e., a sharp corner), which, under two time derivatives, becomes a $\delta$-function. We will thus write
\begin{equation}\label{eq:newK}
    D^{\pm}K^{ij}_{tot}= D^{\pm}K^{ij} + D^{\pm}\bar{K}^{ij} + D^{\pm}K^{ij}_{other},
\end{equation}
where $D^{\pm}K^{ij}$ contains all $\Theta$-function terms, $D^{\pm}\bar{K}^{ij}$ contains all corner terms, and $D^{\pm}K^{ij}_{other}$ contains any other continuous terms that come along for the ride (which, of course, will not contribute any $\delta$-functions, even under the action of second derivatives). To find the Peierls bracket we use Eq.~\eqref{eq:PBdefdisc}, and hence we do not need to find the explicit expression for any continuous terms.

In contrast to the calculation of HRT area flow in Section \ref{sec:HRTflow}, the flow calculation in this section \textit{does} depend on the smoothness of $\tilde K^{ij}$. For the remainder of this section, we will take $\tilde K^{ij}$ to be smooth. This generally suffices to derive the flow only at first order in $\lambda$ around smooth solutions (see the comments at the end of Section \ref{sec:HRTflow}).  Without matter, the extension to all orders turns out to be trivial, since these spacetimes are always locally AdS$_3$, and the kink is just a coordinate artifact. While we expect our result to hold when matter is present, we save a proof of this for future work. At the moment, our results hold only at first order in $\lambda$ when matter is present.

Using our new expression for $K^{ij}$ in Eq.~\eqref{eq:EOMfull} and taking $\lambda$ small, we obtain an equation relating the terms proportional to $\delta(t-t_\Sigma)$,
\begin{equation}\label{eq:EOMfull}
\begin{split}
    0=&-\frac{1}{16\pi G}  \sqrt{h(t,y)}[\partial_t D^{\pm}K^{ij}(t,y)-\partial_t D^{\pm}K(t,y)h^{ij}(t,y)] \\
    &+\beta \sqrt{h(t,y)} \epsilon^{it k}\bigg( \partial^2_t D^{\pm}\bar{K}^j_k(t,y) - 4 \tilde{K}^{lj}(t,y)\partial_tD^{\pm}K_{lk}(t,y) \\
    &- 4 \tilde{K}_{lk}(t,y)\partial_tD^{\pm}K^{lj}(t,y)+  \tilde{K}^j_k(t,y)\partial_tD^{\pm} K(t,y)+ \tilde{K}(t,y) \partial_tD^{\pm}K^j_k(t,y)) \\
    &+\tilde{K}^{lj}(t,y) \partial_t D^{\pm}K_{lk}(t,y) - \tilde{K}_{k}^j(t,y) \partial_{t}D^{\pm}K(t,y) \bigg)\\ 
    &+\beta \sqrt{h(t,y)} \epsilon^{jt k}\bigg( \partial^2_t D^{\pm}\bar{K}^i_k(t,y) - 4 \tilde{K}^{li}(t,y)\partial_tD^{\pm}K_{lk}(t,y) \\
    &- 4 \tilde{K}_{lk}(t,y)\partial_tD^{\pm}K^{li}(t,y)+  \tilde{K}^i_k(t,y)\partial_tD^{\pm} K(t,y)+ \tilde{K}(t,y) \partial_tD^{\pm}K^i_k(t,y)) \\
    &+\tilde{K}^{li}(t,y) \partial_t D^{\pm}K_{lk}(t,y) - \tilde{K}_{k}^i(t,y) \partial_{t}D^{\pm}K(t,y) \bigg)\\
    &+\sqrt{q(\tilde{x}(y))} \delta_{\Sigma}(\gamma,y)\delta(t-t_{\Sigma})\bigg(\frac{1}{8G} q^{AB}(\tilde x(y))\frac{\partial y^i}{\partial \tilde x^A}\frac{\partial y^j}{\partial \tilde x^B}\\
    &-4\pi\beta \perp^k  v^l K_{kl}(y) \perp^i \perp^j -4\pi\beta \perp^k  v^l K_{kl}(y) v^i v^j \\
    &-8\pi\beta v^k  v^l K_{kl}(y) \perp^{(i} v^{j)} + 4\pi\beta q^{AB}(\tilde x(y))\frac{\partial y^i}{\partial \tilde x^A}\frac{\partial y^j}{\partial \tilde x^B}\perp^k v^l K_{kl}(y)\bigg),
\end{split}
\end{equation}
and another equation relating terms proportional to $\delta'(t-t_\Sigma)$,
\begin{equation}
    \partial^2_t(\epsilon^{it k}  D^{\pm}K^j_k(t,y) + \epsilon^{jt k} D^{\pm}K^i_k(t,y)) = 4\pi \frac{\sqrt{q(\tilde{x}(y))}}{\sqrt{h(y)}} \delta_{\Sigma}(\gamma,y) \perp^{(i}  v^{j)} \delta'(t-t_{\Sigma}).
\end{equation}
Integrating the latter equation twice on both sides, we get an equation for the retarded and advanced solutions, respectively:
\begin{eqnarray}\label{eq:retK}
    \epsilon^{it k} D^-K^j_k(t,y) + \epsilon^{jt k} D^-K^i_k(t,y) =& 4\pi \frac{\sqrt{q(\tilde{x}(y))}}{\sqrt{h(y)}} \delta_{\Sigma}(\gamma,y) \perp^{(i}  v^{j)} \Theta(t-t_{\Sigma}), \\
    \label{eq:advK}
    \epsilon^{it k} D^+K^j_k(t,y) + \epsilon^{jt k} D^+K^i_k(t,y) =& -4\pi \frac{\sqrt{q(\tilde{x}(y))}}{\sqrt{h(y)}} \delta_{\Sigma}(\gamma,y) \perp^{(i}  v^{j)} \Theta(t_{\Sigma}-t).
\end{eqnarray}
For now we will work with the retarded solution, saving the advanced solution for later. Contracting Eq.~\eqref{eq:retK} with $\perp_i\perp_j$ and $v_iv_j$, we find
\begin{equation}\label{eq:perpvK}
    \perp^kv_j D^-K^j_k(t,y)=0 \,\,\,\,\text{ and } \,\,\,\, v^k\perp_jD^-K^j_k(t,y)=0.
\end{equation}
To obtain the above equation, we used $\perp_i\epsilon^{itk}=v^k$ and $v_i\epsilon^{itk}=-\perp^k$. To derive these, we note that, due to the normalization and orthogonality constraints of $\perp^i$, $v^i$, and $n^a$, we have
\begin{equation}\label{eq:perpEpDef}
    \perp_i \epsilon^{i0k} = \pm v^k.
\end{equation}
This then gives $\perp_i v_k \epsilon^{i0k} = \pm 1$, and so $\perp_k v_i \epsilon^{i0k} = \mp 1$. We then have
\begin{equation}\label{eq:vEpDef}
    v_i \epsilon^{i0k} = \mp \perp^k.
\end{equation}
In what follows, we choose the plus sign in Eq.~\eqref{eq:perpEpDef} and so we have the minus sign in Eq.~\eqref{eq:vEpDef}. We choose these signs so that our result in the limit $c_L=c_R$ matches what we find for Einstein-Hilbert gravity. This choice of sign appears to be consistent with standard conventions, e.g. in \cite{Carroll_TB}.

Now that we have derived these useful identities, let us contract Eq.~\eqref{eq:retK} with $\perp_iv_j+v_i\perp_j$, yielding
\begin{equation}
    (v_iv_j-\perp_i\perp_j)D^-K^{ij}(t,y) =2\pi \frac{\sqrt{q(\tilde{x}(y))}}{\sqrt{h(y)}} \delta_{\Sigma}(\gamma,y)\Theta(t-t_{\Sigma}).
\end{equation}
These are thus the only components of the solution that survive. We can therefore write the solution as
\begin{equation}\label{eq:DKwithc}
    D^-K^{ij}(t,y) =2\pi \frac{\sqrt{q(\tilde{x}(y))}}{\sqrt{h(y)}} \delta_{\Sigma}(\gamma,y)\Theta(t-t_{\Sigma})(c_vv^iv^j+c_\perp \perp^i\perp^j),
\end{equation}
with $c_v-c_\perp=1$.

We now must solve for $c_v$ and $c_\perp$. We start by contracting Eq.~\eqref{eq:EOMfull} with $\perp_i\perp_j+v_iv_j$. Under this contraction, many terms will cancel out, and we are left with
\begin{equation}
    \begin{split}
        0=&-\frac{1}{16\pi G}\sqrt{h(t,y)}(\perp_i\perp_j+v_iv_j)\partial_t D^{-}K^{ij}(t,y)+ \frac{1}{8\pi G}\sqrt{h(y)}\partial_t D^{-}K(t,y)\\
        &+\beta\sqrt{h(y)}(\perp_i\perp_j+v_iv_j)\epsilon^{itk}K^{lj}(t,y)\partial_t D^-K_{lk}(t,y) \\
        &+\beta\sqrt{h(t,y)}(\perp_i\perp_j+v_iv_j)\epsilon^{jtk}K^{li}(t,y)\partial_t D^-K_{lk}(t,y) \\
        &+\frac{1}{8G}\sqrt{q(\tilde{x}(y))}\delta_\Sigma(\gamma,y)\delta(t-t_\Sigma) - 4\pi\beta\sqrt{q(\tilde{x}(y))}\perp^k v^l K_{kl}(y) \delta_\Sigma(\gamma,y)\delta(t-t_\Sigma).
    \end{split}
\end{equation}
Substituting the right-hand side of Eq.~\eqref{eq:DKwithc} into the equation above, we find
\begin{equation}
        \frac{1}{8 G}(c_\perp+c_v) + 4\pi\beta(c_v-c_\perp)\perp^kv^k K_{kl}(y) = -\frac{1}{8G} + 4\pi\beta\perp^k v^l K_{kl}(y). 
\end{equation}
Since $c_v-c_\perp=1$, the equation above reduces to $c_\perp+c_v=-1$. Then we have $c_\perp=-1$ and $c_v=0$. Using Eq.\eqref{eq:DKwithc}, we obtain the retarded solution
\begin{equation}\label{eq:retKTMG}
\begin{split}
    D^-K^{ij}(t,y) =& - 2\pi \frac{\sqrt{q(\tilde{x}(y))}}{\sqrt{h(y)}} \delta_{\Sigma}(\gamma,y)\Theta(t-t_{\Sigma})\perp^i\perp^j \\
    =& - 2\pi \hat{\delta}_{\Sigma}(\gamma,y)\Theta(t-t_{\Sigma})\perp^i\perp^j.
\end{split}
\end{equation}

To solve for the advanced solution, $D^+K^{ij}$, we start from Eq.~\eqref{eq:advK} and follow the same steps as for the retarded solution. We thus obtain
\begin{equation}\label{eq:advKTMG}
\begin{split}
    D^+K^{ij}(t,y) = 2\pi \hat{\delta}_{\Sigma}(\gamma,y)\Theta(t_{\Sigma}-t)\perp^i\perp^j.
\end{split}
\end{equation}
Combining Equations \eqref{eq:retKTMG} and \eqref{eq:advKTMG} yields the Peierls bracket
\begin{equation}\label{eq:flow}
    \left\{\sigma_{TMG}[R],K^{ij}(t_\Sigma,y)\right\} = -2\pi \hat{\delta}_{\Sigma}(\gamma,y)\perp^i\perp^j.
\end{equation}
This is our main result, and it agrees exactly with Eq.~\eqref{eq:PB_EH}, the result for Einstein-Hilbert gravity. The action of $\sigma_{TMG}[R]$ is thus the same as the HRT area action, generating a boundary-condition-preserving kink transformation as shown in Figure \ref{fig:flow}. In particular, we can write the relation $\sigma_{TMG}[R] = H_R + \mathcal{K}[\gamma]$ as before, where $H_R$ is the generator of the boundary one-sided boost and $\mathcal{K}[\gamma]$ is the kink transform.

\section{Geometric entropy commutators}\label{sec:GEcomms}
In this section, we aim to reproduce the modular commutator result of \cite{Zou_2022}, but from the bulk perspective. We do this in two ways: by using the results above, and by using the boundary stress tensor algebra. We indeed find agreement between our results and \cite{Zou_2022}. In addition, the nature of our calculation allows us to easily extend the results of \cite{Zou_2022} to disjoint boundary regions. This is difficult in their setting as the modular commutator is defined only for three contiguous boundary regions.

We work in topologically massive gravity with negative cosmological constant and without matter, for spacetimes asymptotic to Poincar\'e AdS$_3$. Spacetimes of this form are always diffeomorphic to TMG in Poincar\'e AdS$_3$. Thus, if we start in TMG in vacuum Poincar\'e AdS$_3$, we can obtain any other spacetime of this form via a boundary conformal transformation. Defining $u=t-x$ and $v=t+x$, the Poincar\'e AdS$_3$ metric of Eq.~\eqref{eq:metric} becomes
\begin{equation}
    ds^2 = \frac{1}{z^2}(-dudv+dz^2).
\end{equation}
Then a boundary conformal transformation will be a map $(u,v) \to (U(u),V(v))$, such that the boundary metric becomes
\begin{equation}\label{eq:bndytrans}
    ds_{\partial}^2 = -dUdV = -e^{2\sigma_-(U)}e^{2\sigma_+(V)}dudv.
\end{equation}
We see the flat boundary metric is rescaled by some conformal factor $e^{2\sigma_-(U)}e^{2\sigma_+(V)}$.

We start in Section \ref{sec:T} by deriving the change of the boundary stress energy tensor due to this conformal transformation. Then, in Section \ref{sec:S}, we write the renormalized geometric entropy in this theory. In Section \ref{sec:algebra_flow} we calculate the effect of TMG entropy flow on the stress energy tensor. We then use the geometric flow given by Eq.~\eqref{eq:flow} to calculate, in vacuum Poincar\'e AdS$_3$ and with a specific choice of boundary region $R$, commutators between TMG entropies defined by different boundary regions. In Section \ref{sec:algebra_T} we generalize these results to include general $R$ and all spacetimes diffeomorphic to vacuum Poincar\'e AdS$_3$ (which, in particular, include planar black holes). Finally, in Section \ref{sec:disjoint}, we use our results to understand the entropy algebra with disjoint boundary regions.

\subsection{The boundary stress energy tensor}\label{sec:T}
We write the full boundary stress energy tensor in TMG as $\tilde T_{ij} = T_{ij} - T_{ij}^{CS}$, where $T_{ij}$ is the stress tensor in Einstein-Hilbert gravity and $T_{ij}^{CS}$ represents the contribution from the Chern-Simons term to the stress tensor. If we know the variation for the Chern-Simons term in the action, then $T^{ij}_{CS}$ can be found by evaluating
\begin{equation}\label{eq:stress}
\begin{split}
    \delta I_{CS} =& \frac{1}{2} \int_{\partial \mathcal{M}} T^{ij}_{CS}\delta g_{ij} \sqrt{g} d^2x \\
    =& -\frac{1}{2} \int_{\partial \mathcal{M}} T_{ij}^{CS}\delta g^{ij} \sqrt{g} d^2x.
\end{split}
\end{equation}
We know $\delta I_{CS}$ from Eq.~\eqref{eq:delICS}, however we will instead use the form of the variation found in \cite{Kraus_2006}. Hence, the variation takes the form
\begin{equation}\label{eq:delScs}
\begin{split}
    \delta I_{CS} =& 2\beta \int_{\partial \mathcal{M}} d^2x\sqrt{g} R^{\eta j}_{\,\,\eta k} \delta g_{ij} \epsilon^{i \eta k}  + \beta\int_{\partial \mathcal{M}} d^2x\sqrt{g} [2K^k_i\delta K_{kj}-\Gamma^k_{li}\delta \Gamma^l_{kj}]\epsilon^{ij}\\
    &-\beta\int_{\mathcal{M}} d^3x\sqrt{g} \nabla_{\beta} (R^{\beta \rho}_{\,\,\mu\nu}) \epsilon^{\gamma\mu\nu} \delta g_{\gamma \beta} .
\end{split}
\end{equation} 
In our work below we will use a Fefferman-Graham expansion, writing the metric as
\begin{equation}
    ds^2=d\eta^2 + g_{ij}dx^idx^j,
\end{equation}
where we define $g_{ij}$ as an expansion about the boundary metric $g_{ij}^{(0)}$: $g_{ij}=e^{2\eta}g_{ij}^{(0)} + g_{ij}^{(2)} + ...$. For Poincar\'e AdS in $2+1$-dimensions, we have the Minkowski metric on the boundary, and
\begin{equation}
    g_{ij}^{(2)}=\kappa T_{ij}.
\end{equation}
In vacuum Poincar\'e AdS$_3$ all terms in Eq.~\eqref{eq:delScs} independently vanish, and so $\delta I_{CS}=0$. The first term vanishes since $R^{\eta j}_{\eta k} \propto \delta_k^j$ for large $\eta$, and the last term vanishes since the curvature is covariantly constant. The vanishing of the second term is not as obvious, but it is due to that fact that $g^{(0)}_{ij}=\eta_{ij}$, and $T_{ij}=0$. We thus see that the original $T_{CS}$ (i.e., before the conformal transformation) is zero. 

We will now apply a conformal transformation to Eq.~\eqref{eq:delScs}, then extract the stress energy tensor. Since we are perturbing about a flat boundary metric, all terms vanish except the extrinsic curvature term (we refer the reader to \cite{Kraus_2006} for more details):
\begin{equation}
\begin{split}
    K^k_i\delta K_{kj} \epsilon^{ij}=& -g^{kl}_{(2)}g_{li}^{(0)}\delta g_{kj}^{(0)}\epsilon^{ij} +... \\
    =& \kappa T_{ki} g_{jl}^{(0)} \delta g^{kl}_{(0)}\epsilon^{ij} +... .
\end{split}
\end{equation}
Using Eq.~\eqref{eq:stress}, we can extract $T^{CS}_{ki}$, yielding
\begin{equation}\label{eq:TCSdef}
    T^{CS}_{kl} = -4 \kappa \beta T_{ki}g_{jl}^{(0)} \epsilon^{ij}.
\end{equation}
(This verifies that $T^{CS}_{kl}=0$ when $T_{ki}=0$ in the vacuum.) Now, under the conformal transformation in Eq.~\eqref{eq:bndytrans}, the stress tensor will transform as \cite{Fischetti_2012}
\begin{equation}\label{eq:Ttrans}
    T_{ab}dx^adx^b\to T_{ab}^{original} dx^adx^b + \frac{c_0}{12\pi}[\partial_u^2\sigma-(\partial_u \sigma)^2]du^2 + \frac{c_0}{12\pi}[\partial_v^2\sigma-(\partial_v \sigma)^2]dv^2.
\end{equation} 
Hence, applying this transformation to $T_{ki}$ in Eq.~\eqref{eq:TCSdef}, we get
\begin{equation}
\label{eq:TCSp}
\begin{split}
    T^{CS}_{ab}dx^adx^b \to& -4 \kappa \beta \bigg(\frac{c_0}{12\pi}g_{uv}^{(0)} \epsilon^{uv}[\partial_u^2\sigma-(\partial_u \sigma)^2]du^2 \\
    &\hspace{2.7cm}- \frac{c_0}{12\pi}g_{uv}^{(0)} \epsilon^{uv}[\partial_v^2\sigma-(\partial_v \sigma)^2]dv^2\bigg) \\
    \to& 4 \beta([\partial_u^2\sigma-(\partial_u \sigma)^2]du^2 - [\partial_v^2\sigma-(\partial_v \sigma)^2]dv^2) \\
    \to& 4 \beta([\partial_U^2\sigma+(\partial_U \sigma)^2]dU^2 - [\partial_V^2\sigma+(\partial_V \sigma)^2]dV^2).
\end{split}
\end{equation}
where we used the fact that the original stress tensor is zero. To obtain the second line of the equation, we use the convention $\epsilon^{tx}=-1$ to get
\begin{equation}
     \epsilon^{uv}=\frac{\partial u}{\partial t}\frac{\partial v}{\partial x}\epsilon^{tx}+\frac{\partial u}{\partial x}\frac{\partial v}{\partial t}\epsilon^{xt} = -2.
\end{equation}

Finally, under a conformal transformation from the vacuum, the full stress tensor in this theory becomes
\begin{equation}\label{eq:TtransTMG}
    \begin{split}
        \tilde T_{ij}dx^idx^j \to& \frac{1}{12\pi}(c_0-48\pi\beta) [\partial_U^2\sigma+(\partial_U \sigma)^2]dU^2 \\
        &+ \frac{1}{12\pi}(c_0+48\pi\beta)[\partial_V^2\sigma+(\partial_V \sigma)^2]dV^2 \\
        \to& \frac{c_R}{12\pi} [\partial_U^2\sigma+(\partial_U \sigma)^2]dU^2 + \frac{c_L}{12\pi}[\partial_V^2\sigma+(\partial_V \sigma)^2]dV^2.
    \end{split}
\end{equation}
We define the TMG stress tensor components as $\tilde T_{UU}(U) = \frac{c_L}{12\pi} [\partial_U^2\sigma+(\partial_U \sigma)^2]$ and $\tilde T_{VV}(V) = \frac{c_L}{12\pi} [\partial_V^2\sigma+(\partial_V \sigma)^2]$. By comparison with Eq.~\eqref{eq:Ttrans}, we see that $\tilde T_{UU}(U) = \frac{c_L}{c_0} T_{UU}(U)$ and $\tilde T_{VV}(V) = \frac{c_R}{c_0} T_{VV}(V)$. We note that these relations are more obvious from the CFT perspective, where, under Wick rotation, we can relate the $u$ and $v$ terms of the stress tensor to holomorphic and anti-holomorphic parts $T(z)$ and $T(\bar{z})$, respectively. Then we replace $c_0$ with $c_R$ in $T(z)$ and with $c_L$ in $T(\bar{z})$. Here, however, we wished to understand the stress tensor transformation from the bulk perspective.

\subsection{Geometric entropy in TMG asymptotic to Poincar\'e AdS$_3$}\label{sec:S}
The geometric entropy of TMG is given by Eq.~\eqref{eq:S}, as proven by \cite{Castro_2014} using the replica trick. Now suppose we have a $1+1$D chiral CFT region $R$ anchored at $(u_1,v_1)$ and $(u_2,v_2)$. We take $R$ to be the straight line segment between the anchor points. In vacuum Poincar\'e AdS$_3$, the non-renormalized TMG entropy can be written as \cite{Iqbal_2016}
\begin{equation}\label{eq:TMGareaIW}
    \tilde{\sigma}^{vac}_{TMG}[R] = \frac{c_L}{12}\ln\left( \frac{(v_1-v_2)^2}{\epsilon_{v_1}\epsilon_{v_2}}\right) + \frac{c_R}{12}\ln\left( \frac{(u_1-u_2)^2}{\epsilon_{u_1}\epsilon_{u_2}}\right) 
\end{equation}
where $\epsilon_{u_i}$ and $\epsilon_{v_i}$ for $i=1,2$ are the cut-offs in the $u$ and $v$ directions for each anchor point. To maintain translation invariance we can choose $\epsilon_{u_1}=\epsilon_{u_2}=\epsilon_u$ and $\epsilon_{v_1}=\epsilon_{v_2}=\epsilon_v$. Then, to renormalize the TMG entropy, we follow the standard approach of adding counterterms and taking the limit $\epsilon \to 0$:
\begin{equation}\label{eq:TMGarearenorm1}
\begin{split}
    \sigma^{vac}_{TMG}[R]=& \lim_{\epsilon\to 0}\left(\tilde{\sigma}^{vac}_{TMG}[R] + \frac{c_L}{6}\ln \epsilon_v + \frac{c_R}{6}\ln\epsilon_u\right) \\
    =& \frac{c_L}{6}\ln|v_1-v_2| + \frac{c_R}{6}\ln|u_1-u_2|.
\end{split}
\end{equation}

The renormalized entropy is not invariant under the conformal transformation $(u,v) \to (U(u), V(v))$ given in Eq.~\eqref{eq:bndytrans}. This conformal transformation consists of two parts: a diffeomorphism taking $u \to U(u)$ and $v \to V(v)$, and a Weyl rescaling of the metric. The metric is invariant under such a transformation, but Eq.~\eqref{eq:TMGareaIW} is not, because the cut-offs transform as
\begin{eqnarray}\label{eq:cutoffconf}
    \epsilon_{v_i}& \to e^{2\sigma_+(V_i)}\epsilon_{v_i}, \\
    \epsilon_{u_i}& \to e^{2\sigma_-(U_i)}\epsilon_{u_i}.
\end{eqnarray}
If one wishes to use the same cut-offs before and after the conformal transformation, then the renormalized entropy is defined via the same subtraction, and so transforms by adding $\sigma_\pm$:
\begin{equation}\label{eq:TMGarearenorm}
\begin{split}
    \sigma_{TMG}[R] =& \sigma^{vac}_{TMG}[R] + \frac{c_L}{6} [\sigma_+(V_1) + \sigma_+(V_2)] + \frac{c_R}{6} [\sigma_+(U_1) + \sigma_+(U_2)]. \\
    =& \frac{c_L}{6}\ln|v_2-v_1| + \frac{c_L}{6}\ln|u_1-u_2|  \\
    &+ \frac{c_L}{6} [\sigma_+(V_1) + \sigma_+(V_2)] + \frac{c_R}{6} [\sigma_+(U_1) + \sigma_+(U_2)].
    \end{split}
\end{equation}
This is the renormalized TMG entropy, which is of course similar to the result for the renormalized HRT area under a conformal transformation, except with $c_0$ replaced by $c_L$ in anti-holomorphic terms and by $c_R$ in holomorphic terms. 

In particular, in Poincar\'e AdS$_3$ in Einstein-Hilbert gravity, we instead have
\begin{equation}
    \frac{ A_{HRT}[R]}{4G} = \frac{c_0}{6} (A_V(V_1,V_2) + A_U(U_1,U_2))
\end{equation}
up to a possible constant term which will not factor into our analysis, with 
\begin{eqnarray}
    A_V(V_1, V_2)=&\ln|v(V_2)-v(V_1)| + \sigma_+(V_1)+ \sigma_+(V_2), \\
    A_U(U_1, U_2)=&\ln|u(U_1)-u(U_2)| + \sigma_-(U_1)++ \sigma_-(U_2).
\end{eqnarray}
Hence, we can write the TMG entanglement entropy in terms of the $U$ and $V$ pieces of the HRT-area as as simple rescaling:
\begin{equation}
    \sigma_{TMG}[R] = \frac{c_L}{6} A_V(V_1,V_2) + \frac{c_R}{6} A_U(U_1,U_2).
\end{equation}
In what follows, we will use this expression to rewrite the entropy commutators in \cite{Kaplan_2022}.

\subsection{Entropy algebra from geometric flow}\label{sec:algebra_flow}
We will now use the geometric picture of TMG entropy flow to compute the action of $\sigma_{TMG}[R]$ on the stress tensor, and the commutator between the TMG entropies of two different boundary regions. As discussed above, this entropy flow kinks $\Sigma$ in the bulk, but preserves $\partial \Sigma$. In this section, we work in asymptotically Poincar\'e AdS$_3$ TMG without matter and without black holes, although we will later generalize to spacetimes allowing planar black holes. We follow the same approach as in Section 3 of \cite{Kaplan_2022}, and refer the reader to the discussion there for more details. Our goal here is to review the essential parts of that calculation, and to note any differences (or lack thereof) between TMG and Einstein-Hilbert gravity. 

Our result for the geometric action of the $\sigma_{TMG}$ flow agrees precisely with that for ($1/4G$ times the) HRT-area flow in Einstein-Hilbert gravity. In particular, $\{\sigma[R]/4G, K^{ij}(y)\}$ has not changed with the addition of the Chern-Simons term, and we still have $\sigma[R] = H_R + \mathcal{K}[\gamma]$. The kink transform $\mathcal{K}[\gamma]$ introduces a relative boost between the two sides of $\gamma$, and so leaves $\gamma$ invariant. In asymptotically Poincar\'e AdS$_3$ TMG, the action of $\mathcal{K}[\gamma]$ on the boundary introduces a gravitational anomaly, but this anomaly does not change the equations of motion: the equations of motion change by the addition of the Cotton tensor, which vanishes in Poincar\'e AdS$_D$. Nor does $\mathcal{K}[\gamma]$ change the boundary metric in the boosted wedge. Hence, the action of $\mathcal{K}[\gamma]$ leaves TMG invariant, and we need only consider the action of $H_R$, which must be a boundary conformal transformation. This is the transformation which "undoes" the boundary action of the kink transformation, and so is a boost with a rapidity we will denote as $2\pi\lambda$.

As in \cite{Kaplan_2022}, we take the action of $H_R$ to be a map $(u,v) \to (U(u),V(v))$ defined by \eqref{eq:bndytrans}.\footnote{Note, however, that the purpose of this conformal transformation is different than the purpose of the transformation introduced in Eq.~\eqref{eq:bndytrans}}  We can specify the conformal factor explicitly by taking a boundary region $R_0$, which is the half-line $x \in [0, \infty)$ at $t = 0$ on the boundary at $z = 0$, and considering the extremal surface corresponding to $R_0$. Without matter, this extremal surface is the HRT surface $\gamma_{R_0}$, and it is the bulk geodesic at $x=t=0$ for all $z$. Then
\begin{equation}\label{eq:specific_trans}
    U=ue^{-2\pi\lambda \Theta(-U)},\,\,\,\,V=Ve^{2\pi\lambda \Theta(V)},
\end{equation}
giving
\begin{equation}
    \sigma_-(U) = -\pi\lambda \Theta(-U),\,\,\,\,\sigma_+(V) = \pi\lambda \Theta(V).
\end{equation}
Plugging into Eq.~\eqref{eq:TtransTMG}, the stress tensor under the action of $\sigma_{TMG}[R_0]$ is
\begin{eqnarray}
    T_{VV} =& \frac{c_L}{12\pi}(\lambda \delta'(V) + \pi\lambda^2[\delta(V)]^2) \\
    T_{UU} =& \frac{c_R}{12\pi}(\lambda \delta'(U) + \pi\lambda^2[\delta(U)]^2).
\end{eqnarray}
We can also calculate the effect of $\sigma_{TMG}[R_0]$ on another TMG entropy defined by a different boundary region $R$. This is the same calculation as in Einstein-Hilbert gravity: we write $\sigma_{TMG}[R]$ under the conformal transformation defined in Eq.~\eqref{eq:specific_trans}, thus giving it explicit $\lambda$ dependence. We write this transformed entropy as $\sigma_{TMG,\lambda}[R]$. Then
\begin{equation}
\begin{split}
    \left\{\sigma_{TMG}[R_0], \sigma_{TMG}[R]\right\} =& \frac{d}{d\lambda} \sigma_{TMG,\lambda}[R] \
    \bigg|_{\lambda=0} \\
    =&-\frac{\pi c_L}{3} \frac{V_1\Theta(-V_1V_2)}{V_2-V_1} -\frac{\pi c_R}{3}\frac{U_1\Theta(-U_1 U_2)}{U_1-U_2}.
\end{split}
\end{equation}
In the next section, we generalize this result to spacetimes diffeomorphic to subregions of vacuum Poincar\'e AdS$_3$; in particular, we will now be able to include planar black holes. We also generalize to commutators between TMG entropies defined by arbitrary boundary regions.

\subsection{Entropy algebra from stress tensors}\label{sec:algebra_T}
Now, we consider spacetimes diffeomorphic to subregions of vacuum Poincar\'e AdS$_3$. By allowing for certain singular conformal transformations from the vacuum, i.e. ones where we specify the boundary conditions $v(V=\infty)=0$ and $u(U=\infty)=0$, our solutions asymptote to $M>0$ planar black holes. Otherwise, for solutions that asymptote to Poincar\'e AdS$_3$, we choose $v(V=0)=0$ and $u(U=0)=0$. Let us now proceed with a calculation of the commutator between the geometric entropies of two different boundary regions. We do this by starting from the stress tensor algebra (i.e, the Virasoro algebra with the appropriate chiral central charge), then use the Leibniz rule to get the TMG entropy algebra. We mainly include this section as an independent check on our TMG entropy flow calculation in Section \ref{sec:entropyFlowTMG}. We also include it to make contact with \cite{Kaplan_2022}: in \cite{Kaplan_2022}, we calculated the geometric entropy commutator in Einstein-Hilbert gravity following this same method of starting from the stress tensor algebra. 

The boundary stress tensor algebra in Einstein-Hilbert gravity is \cite{Francesco_TB}
\begin{equation}
\label{eq:sta}
    \{T_{VV}(V),T_{VV}(V')\}= 2 T_{VV}(V')\delta'(V-V')-  T_{VV}'(V')\delta(V-V') - \frac{c_0}{24\pi}\delta^{'''}(V-V'),
\end{equation}
and similarly for the algebra of $T_{UU}$. Suppose we have the boundary region $R$ anchored at $(U_1,V_1)$ and $(U_2,V_2)$, and the boundary region $R'$ anchored at $(U_1',V_1')$ and $(U_2',V_2')$. Without loss of generality, we take $U_1>U_2$, $V_1<V_2$, $U_1'>U_2'$, and $V_1'<V_2'
$. Then, the Leibniz rule is used to obtain the HRT area algebra from the Virasoro algebra:
\begin{equation}\label{eq:SLeibniz}
    \begin{split}
    \bigg\{ \frac{A_{HRT}[R]}{4G}&,\frac{A_{HRT}[R']}{4G}\bigg\} = \\
    &\int dV dV' d\bar{V} d\bar{V}' \frac{1}{4G}\frac{\partial A_{HRT}[R]}{\partial \sigma_+(V)} \frac{\partial \sigma_+(V)}{\partial T_{VV}(V')}\{T_{VV}(V'),T_{VV}(\bar{V}')\} \\
    &\hspace{6cm}\times\frac{\partial \sigma_+(\bar{V})}{\partial T_{VV}(\bar{V}')} \frac{1}{4G}\frac{\partial A_{HRT}[R']}{\partial \sigma_+(\bar{V})} \\
    &+\int dU dU' d\bar{U} d\bar{U}' \frac{1}{4G} \frac{\partial A_{HRT}[R]}{\partial \sigma_-(U)} \frac{\partial \sigma_-(U)}{\partial T_{UU}(U')}\{T_{UU}(U'),T_{UU}(\bar{U}')\} \\
    &\hspace{6.2cm}\times\frac{\partial \sigma_-(\bar{U})}{\partial T_{UU}(\bar{U}')} \frac{1}{4G}\frac{\partial A_{HRT}[R']}{\partial \sigma_-(\bar{U})}.
    \end{split}
\end{equation}

In TMG, using the stress tensor components defined after Eq.~\eqref{eq:TtransTMG}, the Virasoro algebra becomes
\begin{eqnarray}
    \{\tilde T_{VV}(V), \tilde T_{VV}(V')\} =& \frac{c_L}{c_0} \{T_{VV}(V),T_{VV}(V')\}_{E-H} \\
    \{\tilde T_{UU}(U), \tilde T_{UU}(U')\} =& \frac{c_R}{c_0} \{T_{UU}(U),T_{UU}(U')\}_{E-H}.
\end{eqnarray}
where the E-H subscript stands for Einstein-Hilbert. The other terms in the integrals are related to their Einstein-Hilbert counterparts as
\begin{equation}
    \frac{\partial \sigma_{TMG}[R]}{\partial \sigma_+(V)} = \frac{c_L}{4Gc_0} \frac{\partial A_{HRT}[R]}{\partial \sigma_+(V)} \,\,\,\text{ and } \,\,\,\frac{\partial \sigma_{TMG}[R]}{\partial \sigma_-(U)} = \frac{c_R}{4Gc_0} \frac{\partial A_{HRT}[R]}{\partial \sigma_-(U)},
\end{equation}
and
\begin{equation}
    \frac{\partial \sigma_+(V)}{\partial \tilde T_{VV}(V')} = \frac{c_0}{c_L} \frac{\partial \sigma_+(V)}{\partial T_{VV}(V')}\,\,\, \text{ and } \,\,\, \frac{\partial \sigma_-(U)}{\partial \tilde T_{UU}(U')} = \frac{c_0}{c_R} \frac{\partial \sigma_-(U)}{\partial T_{UU}(U')}.
\end{equation}
Putting this all together, Eq.~\eqref{eq:SLeibniz} becomes
\begin{equation}\label{eq:SrelateEH}
    \begin{split}
    \left\{\sigma_{TMG}[R], \sigma_{TMG}[R']\right\} =& \frac{c_L}{c_0}\bigg\{\frac{1}{4G} A_V(V_1,V_2), \frac{1}{4G} A_V(V_1',V_2')\bigg\} \\
    &+ \frac{c_R}{c_0}\bigg\{\frac{1}{4G} A_U(U_1,U_2), \frac{1}{4G} A_U(U_1',U_2')\bigg\}.
    \end{split}
\end{equation}
In Einstein-Hilbert gravity, we had
\begin{eqnarray}
    \bigg\{\frac{1}{4G}A_V(V_1,V_2),\frac{1}{4G}A_V(V_1',V_2')\bigg\} =& \frac{\pi c_0}{6}\begin{cases} 
        2\eta_v - 1, & V_1' <V_1 < V'_2 < V_2 \\
        1 - 2\eta_v , & V_1 < V'_1 <V_2 < V'_2 \\
        0, & \text{otherwise}
    \end{cases} \\
    \bigg\{\frac{1}{4G}A_U(U_1,U_2),\frac{1}{4G}A_U(U_1',U_2')\bigg\} =& \frac{\pi c_0}{6}\begin{cases} 
        2\eta_u -1, & U_2' <U_2 < U'_1 < U_1 \\
        1 - 2\eta_u, & U_2 < U'_2 <U_1 < U'_1 \\
        0, & \text{otherwise},
    \end{cases}
\end{eqnarray}
where we define the cross ratios $\eta_u=\frac{(u_1-u'_1)(u_2-u'_2)}{(u_1-u_2)(u'_1-u'_2)}$ and $\eta_v=\frac{(v_1-v'_1)(v_2-v'_2)}{(v_1-v_2)(v'_1-v'_2)}$. So, plugging these into Eq.~\eqref{eq:SrelateEH}, the entanglement entropy algebra in TMG is given by 
\begin{equation}\label{eq:commutator}
\begin{split}
    \{&\sigma_{TMG}[R],\sigma_{TMG}[R']\} = \\
    &\frac{\pi c_L}{6}\begin{cases} 
        2\eta_v - 1, & V_1' <V_1 < V'_2 < V_2 \\
        1 - 2\eta_v , & V_1 < V'_1 <V_2 < V'_2 \\
        0, & \text{otherwise}
    \end{cases} + \frac{\pi c_R}{6}\begin{cases} 
        2\eta_u -1, & U_2' <U_2 < U'_1 < U_1 \\
        1 - 2\eta_u, & U_2 < U'_2 <U_1 < U'_1 \\
        0, & \text{otherwise}.
    \end{cases} 
\end{split}
\end{equation}
This agrees with the result of \cite{Zou_2022}, given in Eq.~\eqref{eq:Zou}. In vacuum Einstein-Hilbert gravity, the entropy commutator vanishes when we restrict all anchor points to lie on a constant time slice on the boundary. However, in TMG, this configuration instead gives a non-vanishing result. For $x_1'<x_1<x_2'<x_2$,
\begin{equation}
    \{\sigma_{TMG}[R],\sigma_{TMG}[R']\} = \frac{\pi c_-}{6}(2\eta-1),
\end{equation}
with $c_-=c_L-c_R$ and $\eta=\frac{(x_1-x'_1)(x_2-x'_2)}{(x_1-x_2)(x'_1-x'_2)}$. This again agrees with \cite{Zou_2022}.

\subsubsection{Disjoint intervals}\label{sec:disjoint}
\begin{figure}
    \centering
    \begin{tikzpicture}[scale=3]
    \draw (-1,-1) -- (-.5,1) -- (.3,0) -- (-.2,-2) -- (-1,-1);

    \draw (-1.1,.9) node {Planar boundary};
    \draw[->] (-.15,-2) -- (-.05,-1.6) node [below right] {$U$};
    \draw[->] (-.25,-2) -- (-.5,-1.69) node [below left] {$V$};

    \draw[red,thick] (-.85,-.9) -- (-.3,-.75) node [above, midway] {$A$};
    \draw[blue,thick] (-.3,-.75) -- (-.15,-.45);
    \draw[blue,thick] (-.29,-.57) node {$B$};
    \draw[yellow,thick] (-.15,-.45) -- (.1,-.2) node [above=.05cm, midway] {$C$};

    \draw[teal, thick] (-.3,-.75) arc[start angle=-90,delta angle=145,x radius=7mm, y radius=3.03mm];
    \draw[teal] (.4,-.29) node {$\gamma_2$};
    \fill[teal] (-.3,-.75) circle (.4pt);
    \fill[teal] (.1,-.2) circle (.4pt);
    
    \draw[violet,thick] (-.85,-.9) arc[start angle=-90,delta angle=152.2,x radius=15mm, y radius=2.4mm];
    \draw[violet] (.6,-.52) node {$\gamma_1$};
    \fill[violet] (-.85,-.9) circle (.4pt);
    \fill[violet] (-.15,-.45) circle (.4pt);

\end{tikzpicture}
    \caption{For contiguous CFT regions $A$, $B$, and $C$, we can draw $\gamma_1$, the extremal surface corresponding to region $AB$, and $\gamma_2$, the extremal surface corresponding to region $BC$. This is the configuration studied in \cite{Zou_2022}, where the authors find the modular commutator $J(A,B,C)_{\Omega}$, equivalent to the commutator between the TMG entropies of $AB$ and $BC$.}
    \label{fig:contiguous}
\end{figure}
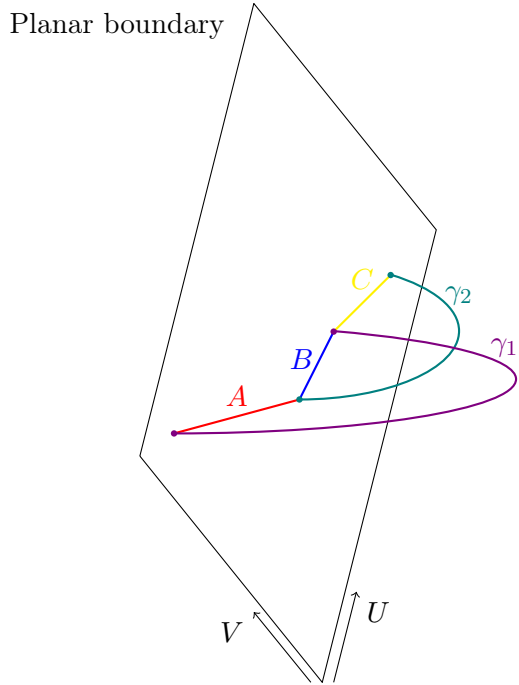

\begin{figure}
    \centering
    \begin{tikzpicture}[scale=2.7]
    \draw (-1,-1) -- (-.5,1) -- (.3,0) -- (-.2,-2) -- (-1,-1);

    \draw (-1.1,.9) node {Planar boundary};
    \draw[->] (-.15,-2) -- (-.05,-1.6) node [below right] {$U$};
    \draw[->] (-.25,-2) -- (-.5,-1.69) node [below left] {$V$};

    \draw[red,thick] (-.85,-.9) -- (-.3,-.75) node [above, midway] {$A$};
    \draw[blue,thick] (-.3,-.75) -- (-.15,-.45);
    \draw[blue,thick] (-.29,-.57) node {$B$};
    \draw[yellow,thick] (-.05,-.25) -- (.2,0) node [above=.05cm, midway] {$C$};
    \draw[green, thick] (-.15,-.45) -- (-.05,-.25);
    \draw[green,thick] (-.17,-.32) node {$D$};

    \draw[violet,thick] (-.85,-.9) arc[start angle=-90,delta angle=152.2,x radius=15mm, y radius=2.4mm];
    \draw[violet] (.6,-.52) node {$\gamma_1$};
    \fill[violet] (-.85,-.9) circle (.4pt);
    \fill[violet] (-.15,-.45) circle (.4pt);

    \draw[teal, thick] (-.3,-.75) arc[start angle=-90,delta angle=163,x radius=5mm, y radius=1.52mm];
    \fill[teal] (-.3,-.75) circle (.4pt);
    \fill[teal] (-.15,-.45) circle (.4pt);

    \draw[teal, thick] (-.05,-.25) arc[start angle=-90,delta angle=162,x radius=8mm, y radius=1.29mm];
    \fill[teal] (-.05,-.25) circle (.4pt);
    \fill[teal] (.2,0) circle (.4pt);

    \draw[thick, ->] (.37,-.43) --(.22,-.55);
    \draw[teal] (.42,-.39) node {$\gamma_2$};
    \draw[thick, ->] (.47,-.35) -- (.62,-.23);
    
\begin{scope}[xshift=2.8cm]
    \draw (-1,-1) -- (-.5,1) -- (.3,0) -- (-.2,-2) -- (-1,-1);

    \draw (-1.1,.9) node {Planar boundary};
    \draw[->] (-.15,-2) -- (-.05,-1.6) node [below right] {$U$};
    \draw[->] (-.25,-2) -- (-.5,-1.69) node [below left] {$V$};

    \draw[red,thick] (-.85,-.9) -- (-.3,-.75) node [above, midway] {$A$};
    \draw[blue,thick] (-.3,-.75) -- (-.15,-.45);
    \draw[blue,thick] (-.29,-.57) node {$B$};
    \draw[yellow,thick] (-.05,-.25) -- (.2,0) node [above=.05cm, midway] {$C$};
    \draw[green, thick] (-.15,-.45) -- (-.05,-.25);
    \draw[green,thick] (-.17,-.32) node {$D$};
    
    \draw[violet,thick] (-.85,-.9) arc[start angle=-90,delta angle=152.2,x radius=15mm, y radius=2.4mm];
    \draw[violet] (.5,-.49) node {$\gamma_1$};
    \fill[violet] (-.85,-.9) circle (.4pt);
    \fill[violet] (-.15,-.45) circle (.4pt);

    \draw[teal, thick] (-.3,-.75) arc[start angle=-90,delta angle=158,x radius=13mm, y radius=3.9mm];
    \fill[teal] (-.3,-.75) circle (.4pt);
    \fill[teal] (.2,0) circle (.4pt);

    \draw[teal, thick] (-.15,-.45) arc[start angle=-90,delta angle=158,x radius=3mm, y radius=1.01mm];
    \fill[teal] (-.15,-.45) circle (.4pt);
    \fill[teal] (-.05,-.25) circle (.4pt);

    \draw[thick, ->] (.35,-.25) -- (.15,-.3);
    \draw[teal] (.42,-.25) node {$\gamma_2$};
    \draw[thick, ->] (.49,-.23) -- (.69,-.18);
\end{scope}

\end{tikzpicture}
    \caption{For contiguous CFT regions $A$ and $B$, and disconnected region $C$, we can draw $\gamma_1$, the extremal surface corresponding to region $AB$, and $\gamma_2$, the extremal surface corresponding to region $BC$. We label the region between $B$ and $C$ as region $D$. As opposed to the contiguous case, $\gamma_2$ splits into two surfaces. In the left figure, $\gamma_2$ is the HRT surface corresponding to region $B$ combined with the HRT surface corresponding to $C$. In the right figure, $\gamma_2$ is the HRT surface corresponding to region $BDC$ combined with the HRT surface corresponding to $D$. This is a configuration we can now study using Eq.~\eqref{eq:commutator}.}
    \label{fig:disjoint}
\end{figure}
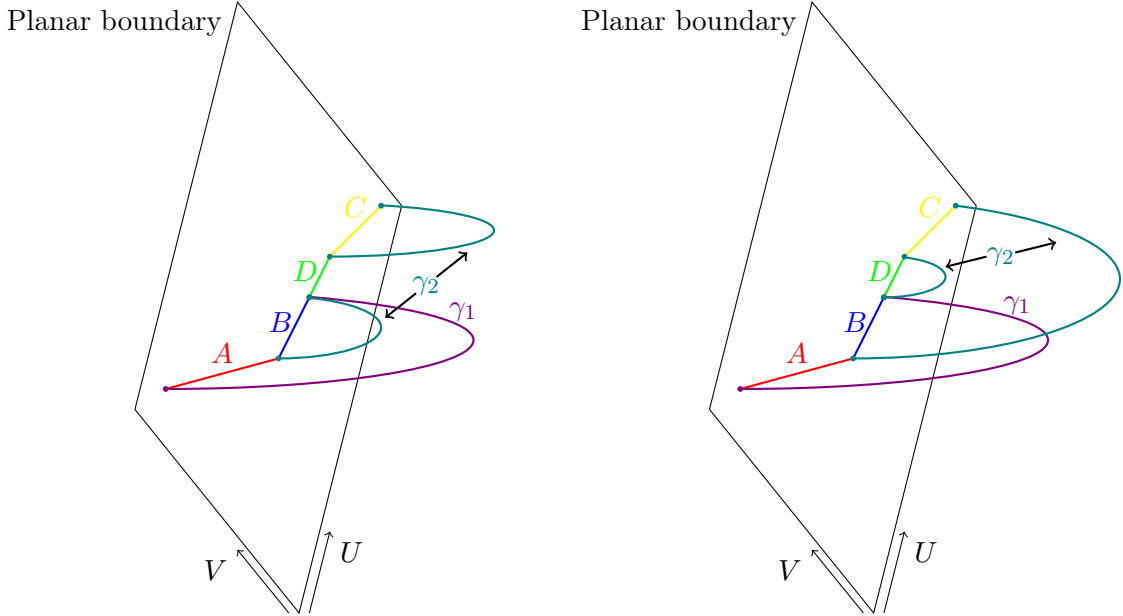

We can also apply our results to slightly more general situations than those considered in \cite{Zou_2022}. In particular, in the semiclassical approximation, their result calculates the commutator between $\sigma_{TMG}[AB]$ and $\sigma_{TMG}[BC]$, the geometric entropies of boundary regions $AB$ and $BC$, respectively, where $A$, $B$ and $C$ are contiguous. See Figure \ref{fig:contiguous} for an illustration. By contrast, the commutators calculated in this work can be defined for disjoint intervals $A$, $B$ and $C$. This is because, in Eq.~\eqref{eq:S}, the TMG entropy is defined as an integral over the extremal surface. Hence, if we have a disconnected surface, the integral splits into two, and the contributions from each piece are additive.

For instance, take $B$ and $C$ to be disjoint. We take the anchor points of $A$ to be $(U_1,V_1)$ and $(U_2,V_2)$, the anchor points of $B$ to be $(U_2,V_2)$ and $(U_3,V_3)$, and the anchor points of $C$ to be $(U_4,V_4)$ and $(U_5,V_5)$. Additionally, we will define a new region $D$ between $B$ and $C$, that is anchored at $(U_3,V_3)$ and $(U_4,V_4)$. Then the bulk extremal surfaces corresponding to boundary region $BC$ have two possible configurations, as shown in Figure \ref{fig:disjoint}. Thus, $\sigma_{TMG}[BC]$ is given by the configuration with minimal entropy:
\begin{equation}
    \begin{split}
        \sigma_{TMG}[BC] = \min \bigg[& \sigma_{TMG}[B]+ \sigma_{TMG}[C], \sigma_{TMG}[BDC] + \sigma_{TMG}[D]\bigg].
    \end{split}
\end{equation}
We can hence define $\sigma_{TMG}[BC]$ in terms of TMG entropies defined by contiguous boundary regions, and thus compute $\sigma_{TMG}$ commutators.

In particular, in the disconnected phase (the left diagram in Fig.~\ref{fig:disjoint}), both $\sigma_{TMG}[B]$ and  $\sigma_{TMG}[C]$ commute separately with the $\sigma_{TMG}[AB]$. Then,
\begin{equation}
    \{\sigma_{TMG}[AB],\sigma_{TMG}[BC]\}_{disconnected}=0.
\end{equation}
In the connected phase (the right diagram of Fig.~\ref{fig:disjoint}), we see that $\sigma_{TMG}[D]$ commutes with $\sigma_{TMG}[AB]$, but $\sigma_{TMG}[BCD]$ and $\sigma_{TMG}[AB]$ do not commute. Without loss of generality, we choose $U_1>U_3$, $V_1<V_3$, $U_2>U_5$, and $V_2< V_5$. This yields
\begin{equation}
\begin{split}
    \{&\sigma_{TMG}[AB],\sigma_{TMG}[BC]\}_{connected} = \\
    &\frac{\pi c_L}{6}\begin{cases} 
        2\eta_v - 1, & V_2 <V_1 < V_5 < V_3 \\
        1 - 2\eta_v , & V_1 < V_2 <V_3 < V_5 \\
        0, & \text{otherwise}
    \end{cases} + \frac{\pi c_R}{6}\begin{cases} 
        2\eta_u -1, & U_5 <U_3 < U_2 < U_1 \\
        1 - 2\eta_u, & U_3 < U_5 <U_1 < U_2 \\
        0, & \text{otherwise}.
    \end{cases} 
\end{split}
\end{equation}
Thus, we have a generalization of Eq.~\eqref{eq:Zou} to disjoint boundary intervals.

\section{Discussion}\label{sec:discussion}
This work began by studying the flow on the covariant phase space induced by geometric entropy in topologically massive gravity, computed in spacetimes asymptotic to AdS$_3$ with standard matter. In terms of Cauchy data on a Cauchy slice $\Sigma$ containing the HRT surface, we found exactly the same result as in \cite{Kaplan_2022} for HRT area flow in Einstein-Hilbert gravity. In particular, the flow leaves the induced metric invariant but shifts the extrinsic curvature by a $\delta$-function as described by Eq.~\eqref{eq:flow}, essentially boosting the entanglement wedge of $R$ relative to that of the complementary region. Without matter, this result holds to all orders in the flow parameter $\lambda$; with matter, our result holds only to first order in $\lambda$. We save the generalization to finite $\lambda$ for future work.

After deriving the geometric entropy flow, we used it to explicitly compute the commutator between TMG entropies defined by different boundary CFT regions. We also derived this commutator by extrapolating from the stress tensor algebra. Our commutators agree with the modular commutator found in \cite{Zou_2022}, the original motivation for this work. We concluded with a short discussion about applying our results to disjoint boundary regions, which is difficult to do with the modular commutator.

It is perhaps surprising that geometric entropy flow in TMG is precisely the same as HRT area flow in Einstein-Hilbert gravity. Arriving at this result through the Peierls bracket method was rather complicated, and required many cancellations between terms. This suggests there may be a more elegant way to approach this calculation, which would make the physical mechanisms behind these cancellations more obvious. Understanding this result more fully could allow generalizations to higher dimensional theories with boundary chiral CFTs, and potentially to theories with other types of higher derivative terms. Geometric entropy flow in higher derivative theories of gravity will be explored in \cite{Dong_Unpublished_2}. Our work here is an important first step to understanding geometric entropy flow more generally.

In the same vein, our Peierls bracket calculation could be extended to higher dimensional theories with boundary chiral CFTs. Explicit formulas for the corresponding geometric entropies have been computed in, for example, $3+1$, $4+1$, and $6+1$ bulk dimensions \cite{Azeyanagi_2015, Ali_2017}. The Peierls bracket calculation for geometric entropy flow in higher dimensions would then follow the same steps as in our work here (expect would be considerably more complicated). As already mentioned, it would thus be helpful to have a more elegant understanding of our result instead of resorting to an explicit calculation.

It would also be interesting to compute explicit TMG entropy commutators in more general configurations, e.g. with matter present or in higher dimensions. Indeed, this has not yet been done for HRT area commutators in Einstein-Hilbert gravity. Additionally, area commutators may have further implications tensor network models, as will be explored in the forthcoming work \cite{Held_Unpublished_1,Held_Unpublished_2}. We would also like to understand the implications of our TMG entropy commutator on tensor network constructions, especially since TMG is an example of a higher derivative theory.

\acknowledgments

The author would like to give special thanks to her advisor, Donald Marolf, without whom this work would not have been possible. She would also like to thank Xi Dong, Jesse Held, Pratik Rath, Arvin Shahbazi-Moghaddam, Jon Sorce, Jie-qiang Wu, and Yijian Zou for many useful discussions on this work. This material is based upon work supported by the Air Force Office of Scientific Research under award number FA9550-19-1-0360, and by funds from the University of California.

\bibliographystyle{jhep}
	\cleardoublepage

\renewcommand*{\bibname}{References}

\bibliography{EEaction}

 \end{document}